\documentclass[a4paper,11pt]{article}
\pdfoutput=1 

\usepackage{jcappub} 

\usepackage[T1]{fontenc} 
\usepackage{times}
\usepackage[shortlabels]{enumitem}

\usepackage[utf8]{inputenc}      
\usepackage[usenames,dvipsnames]{xcolor}

\newcommand{\mr}[1]{\mathrm{#1}}
\newcommand{\mb}[1]{\mathbf{#1}}
\newcommand{\M}[1]{\mathsf{#1}}

\title{\boldmath Largest scales from the largest galaxy surveys: the pseudo Karhunen-Lo\`{e}ve method}

\author{H. S. Xavier}
\affiliation{Instituto de Astronomia, Geof\'{i}sica e Ci\^{e}ncias Atmosf\'{e}ricas da Universidade de S\~{a}o Paulo,\\
Rua do Mat\~{a}o, 1226, Cidade Universit\'{a}ria, S\~{a}o Paulo, SP, Brasil}

\emailAdd{hsxavier@if.usp.br}

\abstract{The increasing area and depth of galaxy surveys will give us access to the largest scales in the Universe and allow 
for a direct test of the primordial power spectrum set by inflation. 
To take full advantage of the survey's volume, we must 
deal with redshift space distortions, growth of structure along the line of sight, luminosity-dependent bias, wide-angle effects 
and complex galaxy selection functions. We present a thorough description of the pseudo Karhunen-Loève method for measuring 
galaxy clustering, a method particularly well-tuned for the largest scales, and extend its applicability and power by taking 
into account light-cone effects, galaxy bias evolution, and by generalizing it to anisotropic selection functions. 
We also show that the combination of non-overlapping surveys result in more information than the sum of its parts 
  and that clustering amplitude evolution along the line of sight, both due to galaxy bias and structure growth, 
  must be taken into account at scales beyond the turn-over.}

\begin{document}
\maketitle
\flushbottom

\section{Introduction}
\label{sec:intro}

Galaxy surveys provide a wealth of information about the Universe 
\cite{Percival01,Eisenstein05,Blake11b,Ross13,Sorensen12,Sanchez13,Laurent16,Bengaly17,Novaes18,Carvalho18} 
and different methods for analyzing galaxy clustering have been proposed in the literature 
\cite{Jeong10,Leistedt12}, focusing on: minimizing the variance and bias of 2-point statistics estimators
\cite{LandySzalay93,Feldman94}; minimizing the mode-coupling effect caused by 
the finite survey volume or partial sky coverage \cite{Tegmark95}; dealing with redshift space distortions (RSD)
\cite{Heavens95,Hamilton98,Yamamoto06,Kazin12}, luminosity-dependent bias \cite{Percival04}, 
growth and non-linear structure \cite{Eisenstein07,Simpson13}, multiplicative errors in the 
selection function \cite{Shafer15,Xavier19} and contamination by stars and quasars \cite{Elsner17}. 

One such method is the so-called pseudo Karhunen-Lo\`{e}ve (pKL) method \cite{Tegmark97,Tegmark98}, 
first introduced in the context of cosmology in \cite{Vogeley96} and applied to the Two-degree-Field 
galaxy redshift survey (2dF) \cite{Colless01,Tegmark02} and to the Sloan Digital Sky Survey 
(SDSS) \cite{York00,Tegmark04}. The pKL method is based on the Karhunen-Lo\`{e}ve transform, 
a statistical method closely related to Principal Component Analysis in which a stochastic signal 
is decomposed into a linear combination of orthogonal functions $\Psi_i(\mb{r})$ tailored to carry as much information 
as possible from that particular signal, according to its expected noise and signal covariance matrices. 
In its application to cosmology, this method is also accompanied by: the orthogonalization of the 
basis functions with respect to templates of potential systematics (which may represent 
contamination, uncertainties in the selection function and other unwanted 
physical effects); and the use of a spherical basis around the observer that allows for the modeling 
of line-of-sight effects such as RSD, photometric redshift (photo-$z$) errors, 
luminosity dependent bias, light-cone constraints and galaxy bias evolution in large regions of the sky, 
where the plane-parallel approximation cannot be applied. Finally, the pKL method includes a simplification 
responsible for the ``pseudo'' qualification: the functions $\Psi_i(\mb{r})$ are assumed to be 
separable into radial and angular parts, $\Psi_i(\mb{r}) = \Psi^{\mr{z}}_i(r)\Psi^{\mr{\theta}}_i(\mb{\hat{r}})$, 
where the angular functions are chosen independently from the radial part (although the converse is not true). 

The pKL method has several advantages over other approaches for studying galaxy clustering, 
particularly for surveys large in area and in depth like Euclid \cite{Lumb09}, LSST \cite{LSST09}, 
DESI \cite{DESI16} and J-PAS \cite{Benitez14}. As mentioned above, the use of a spherical basis grants easy and 
precise modeling of several important effects. 
Since the speed of light is finite and distant observations also probe farther in the past, 
deep surveys have to take into account galaxy bias evolution, growth of structure and other 
light-cone constraints. Moreover, observational effects all follow spherical symmetry, such 
that deep and wide surveys display a geometry that resembles a cone and not a box. Finally,       
the pKL method of orthogonalizing the measurements with respect to systematics is more 
robust than other alternatives such as mode projection \cite{Elsner17} since the latter 
depends on an estimate of the systematics contribution level, while the former does not.

As an optimal analysis method for the largest scales, the pKL method can be valuable 
in the assessment of various hypotheses. For instance, it was shown that primordial 
non-Gaussianities set by certain inflationary models lead to scale-dependent halo biases 
that manifest themselves on Fourier modes with wavenumbers $k\lesssim0.01 h \mr{Mpc^{-1}}$ \cite{Dalal08}, 
and that a strong constraint can be placed on the respective parameter 
$f_{\mr{NL}}$ by large galaxy surveys \cite{Putter17}. Likewise, it has been shown that 
if galaxy formation happens only in regions where matter density contrast reaches a certain threshold 
(threshold biasing), then galaxy bias would also be scale-dependent at similar scales
\cite{Durrer03}. Large galaxy surveys can also be used to test for dark energy clustering, 
which would add power to scales greater then its sound horizon \cite{Takada06b}. Finally, 
probing scales $k\lesssim 0.02 h \mr{Mpc^{-1}}$ (beyond the turn-over of the power spectrum) with sufficient 
precision would provide a direct measurement of the epoch of matter-radiation equality that is independent 
from CMB observations, serving as an interesting consistency check.

In this paper we provide a comprehensive description of the pKL method,  
covering gaps in the literature, and extend its applications to deep and wide surveys 
by including all light-cone and radial effects presented above and by allowing for 
its application to non-separable selection functions that are piece-wise separable. As demonstrated by SDSS 
\cite{Ross12b}, even well calibrated surveys might suffer from selection effects that 
make the radial selection function dependent on the direction of the line of sight; and surveys relying on different  
instruments for different regions of the sky (e.g. the Euclid ground 
segment\footnote{\url{http://www.euclid-ec.org/?page_id=2625}} \cite{Jain15}) may 
also require spatially varying selection functions \cite{Bilicki14,Xavier19}. As a bonus, our treatment permits 
the combination of multiple surveys, and such combination was shown to significantly 
improve their constraints on the amplitude of matter perturbations $\sigma_8$ and the total mass of 
neutrinos $\sum m_\nu$ \cite{Jain15}. As we show below, the pKL method is particularly useful to measure 
large scale perturbations and features that do not depend on the late-time growth of structure, 
such as the transfer function and the primordial power spectrum set by inflation. 

This paper is organized as follows: in Sec. \ref{sec:kl-method} we describe in detail and in 
general terms the properties of the Karhunen-Lo\`{e}ve (KL) modes -- orthogonalized with respect to 
systematics -- and how they are built in practice. In Sec. \ref{sec:pseudo-kl} we present the process of 
building pseudo KL modes, that is, building the modes' angular part independently from the radial part. 
Sec. \ref{sec:pkl-ang} shows a method for optimizing the angular part while Sec. \ref{sec:pkl-radial} 
demonstrates how the radial part is built taking into account evolving bias, structure growth and RSD, 
all in the light-cone. In Sec. \ref{sec:nonsep-pKL} we expand the applicability of the pKL method to 
non-separable (but piece-wise) selection functions, and Sec. \ref{sec:apply-pkl} shows how the pKL modes 
are used to measure the power spectrum, emphasizing their appropriateness for non-evolving features 
and the importance of taking light-cone effects into account. 
We present our final remarks in Sec. \ref{sec:discussion}. 

\section{The Karhunen-Lo\`{e}ve method with orthogonalization to systematics}
\label{sec:kl-method}

\subsection{Describing the data}

Let us define the observed number density of sources classified as galaxies $n_{\mr{obs}}(\mb{r})$ 
at position $\mb{r}$ in redshift space as:

\begin{equation}
n_{\mr{obs}}(\mb{r}) = W(\mb{r})\{ \bar{n}_{\mr{g}}(\mb{r})[1+\delta_{\mr{g}}(\mb{r})] + \epsilon(\mb{r}) + s(\mb{r}) \}.
\label{eq:nobs-def}
\end{equation}
In the equation above, $W(\mb{r})$ is the survey window function with values either 1 or 0, describing 
the limits of the survey; $\bar{n}_{\mr{g}}(\mb{r})$ is the so-called selection function, which gives the
expected number density in a completely homogeneous universe 
and in the absence of contaminations; $\delta_{\mr{g}}(\mb{r})$ is the galaxy density contrast, with 
$\langle \delta_{\mr{g}}(\mb{r}) \rangle = 0$; 
$\epsilon(\mb{r})$ are Poisson fluctuations in the observed density of galaxies; 
and $s(\mb{r})$ are systematics that might include contamination by stars and the dipole due to our own 
peculiar velocity with respect to the Cosmic Microwave Background (CMB) rest-frame \cite{Hamilton98}. 
The position $\mb{r}$ is written $\mb{r}=r\mb{\hat{r}}$, where $\mb{\hat{r}}$ is an unit vector describing the direction 
of $\mb{r}$ (i.e. the angular position) and $r$ is the comoving distance corresponding to the observed redshift $z$, which may 
include the contribution from peculiar velocities \cite{Hamilton98}.

The systematics are considered to be non-statistical in nature, e.g. 
$\langle s(\mb{r}) s(\mb{r'}) \rangle = \langle s(\mb{r}) \rangle \langle s(\mb{r'}) \rangle = s(\mb{r}) s(\mb{r'})$. 
The Poisson fluctuations are 
assumed to have zero mean ($\langle \epsilon(\mb{r}) \rangle = 0$) and to be independent from the signal and from the systematics, 
$\langle \bar{n}_{\mr{g}}(\mb{r})[1+\delta_{\mr{g}}(\mb{r})] \epsilon(\mb{r}) \rangle = \langle s(\mb{r}) \epsilon(\mb{r}) \rangle = 0$. 
Its variance is given by:

\begin{equation}
\langle \epsilon(\mb{r})\epsilon(\mb{r'}) \rangle = \bar{n}_{\mr{obs}}(\mb{r}) \delta_{\mr{D}}^3(\mb{r}-\mb{r'}) = 
[\bar{n}_{\mr{g}}(\mb{r}) + s(\mb{r})]  \delta_{\mr{D}}^3(\mb{r}-\mb{r'}),
\label{eq:poisson-var}
\end{equation}
where $\delta_{\mr{D}}^3(\mb{r}-\mb{r'})$ is the 3D Dirac delta function.

\subsection{Desired properties}
\label{sec:kl-properties}

Our plan is to build a set of mode functions $\Psi_i(\mb{r})$ such that the maximum amount of 
clean cosmological information in $n_{\mr{obs}}(\mb{r})$ [that is, $\delta_{\mr{g}}(\mb{r})$] 
is encoded in the coefficients $x_i$. In other words, $x_i$ is a proxy for the Fourier 
transform of $\delta_{\mr{g}}(\mb{r})$, generalized to optimize to a specific survey geometry; thus, 
$\langle x_i x^*_j \rangle$ becomes our generalized estimator for the power spectrum. The coefficients 
are given by: 

\begin{equation}
x_i \equiv \int \frac{n_{\mr{obs}}(\mb{r})}{w(\mb{r})} \Psi_i(\mb{r}) \mr{d^3}r,
\label{eq:pkl-coeff}
\end{equation} 
where $w(\mb{r})$ are arbitrary (inverse) weights and the integral is performed over the whole 3D space; in 
\cite{Tegmark04}, for instance, $w(\mb{r}) = \bar{n}_{\mr{g}}(\mb{r})$. We now list three 
important properties that we want our mode functions to have:

\begin{enumerate}
\item
We want $\Psi_i(\mb{r})$ such that $x_i$ is insensitive to the contents in 
Eq. \ref{eq:nobs-def} that are void of cosmological information. This can be done by making 
$\Psi_i(\mb{r})$ orthogonal to the systematics $s(\mb{r})$ and to the selection function 
$\bar{n}_{\mr{g}}(\mb{r})$, both weighted by $w(\mb{r})$ and inside $W(\mb{r})$. Given that 
$\langle n_{\mr{obs}}(\mb{r}) \rangle = \bar{n}_{\mr{g}}(\mb{r}) + s(\mb{r})$, this requires 
\cite{Tegmark98}:

\begin{equation}
\langle x_i \rangle = \int \frac{\langle n_{\mr{obs}}(\mb{r}) \rangle}{w(\mb{r})} \Psi_i(\mb{r}) \mr{d^3}r = 0.
\label{eq:x-mean}
\end{equation} 
Let us describe $\langle n_{\mr{obs}}(\mb{r}) \rangle$ as a linear combination of $N$ 
components $M_j(\mb{r})$ with arbitrary amplitudes $a_j$ [so both $\bar{n}_{\mr{g}}(\mb{r})$ and
$s(\mb{r})$ might include multiple contributions], 
$\langle n_{\mr{obs}}(\mb{r}) \rangle = \sum_j^N a_jM_j(\mb{r})$. If we enforce the condition

\begin{equation}
\int \frac{M_j(\mb{r})}{w(\mb{r})} \Psi_i(\mb{r}) \mr{d^3}r = 0,
\label{eq:m-orthogonal}
\end{equation} 
then Eq. \ref{eq:x-mean} is satisfied for \emph{any} value of $a_j$, 
which is a special advantage of the pKL method. Note that the templates $M_j(\mb{r})$ 
must of course include $W(\mb{r})$, and their linear combination must result in 
$\langle n_{\mr{obs}}(\mb{r}) \rangle$. Also, a wrong estimate of $\bar{n}_{\mr{g}}(\mb{r})$ 
may still bias $\langle x_i x^*_j \rangle$ given its multiplicative effect on $\delta_{\mr{g}}(\mb{r})$, 
an issue that plagues most estimators.

Since we set $\langle x_i \rangle = 0$ above, the covariance matrix 
of $x_i$ is simply $\langle x_i x^*_j \rangle$:

\begin{align}
\langle x_i x^*_j \rangle &= 
\int \frac{\langle n_{\mr{obs}}(\mb{r}) n_{\mr{obs}}(\mb{r'})\rangle}{w(\mb{r})w(\mb{r'})} 
\Psi_i(\mb{r}) \Psi^*_j(\mb{r'}) \mr{d^3}r \mr{d^3}r' = S_{ij} + N_{ij}, \label{eq:x-cov} \\
S_{ij} &\equiv \int W(\mb{r})W(\mb{r'})\frac{\bar{n}_{\mr{g}}(\mb{r})\bar{n}_{\mr{g}}(\mb{r'})}{w(\mb{r})w(\mb{r'})}
\langle \delta_{\mr{g}}(\mb{r})\delta_{\mr{g}}(\mb{r'}) \rangle \Psi_i(\mb{r}) \Psi^*_j(\mb{r'}) \mr{d^3}r \mr{d^3}r', 
\label{eq:s-cov}\\
N_{ij} &\equiv \int W(\mb{r})W(\mb{r'})\frac{\langle \epsilon(\mb{r})\epsilon(\mb{r'}) \rangle}{w(\mb{r})w(\mb{r'})}
\Psi_i(\mb{r}) \Psi^*_j(\mb{r'}) \mr{d^3}r \mr{d^3}r', \label{eq:n-cov-0}
\end{align} 
where $\M{S}$ and $\M{N}$ are the signal and noise covariance matrices with elements $S_{ij}$ and $N_{ij}$,
respectively, and the remaining terms proportional to $\bar{n}_{\mr{g}}(\mb{r})$ and $s(\mb{r})$ were eliminated by Eq. 
\ref{eq:m-orthogonal}. Eqs. \ref{eq:poisson-var} and \ref{eq:n-cov-0} boil down to:
 
\begin{equation}
N_{ij} = \int W(\mb{r})\frac{\bar{n}_{\mr{obs}}(\mb{r})}{w^2(\mb{r})}
\Psi_i(\mb{r}) \Psi^*_j(\mb{r}) \mr{d^3}r.
\label{eq:n-cov}
\end{equation} 

\item The covariance matrix $\langle x_i x^*_j \rangle$ has only terms proportional to 
$\langle \delta_{\mr{g}}(\mb{r})\delta_{\mr{g}}(\mb{r'}) \rangle$ and $\langle \epsilon(\mb{r})\epsilon(\mb{r'}) \rangle$, 
and what we want is to maximize the first over the second. To put in more conventional terms, 
we can reach this by first having $\Psi_i(\mb{r})$ such that $N_{ij}=\delta^{\mr{K}}_{ij}$ is an identity matrix 
($\delta^{\mr{K}}_{ij}$ is the Kronecker delta).

\item Since the noise has been normalized to one in the item above, our goal of maximizing signal 
over noise per mode is finally achieved by having $\Psi_i(\mb{r})$ that also diagonalizes $S_{ij}$ (so the 
signal is concentrated in single modes and not dispersed through correlations among modes) and 
that selects the modes with highest signal variance. 
\end{enumerate}

\subsection{Obtaining the desired properties in practice}
\label{sec:kl-mechanics}

The actual process for achieving the three desired properties described in Sec. \ref{sec:kl-properties} 
involves transforming our problem into a linear algebra problem by first binning or band-limiting 
the data $n_{\mr{obs}}(\mb{r})$ by integrating it through a set of basis functions $\Phi_i(\mb{r})$:

\begin{equation}
y_i \equiv \int \frac{n_{\mr{obs}}(\mb{r})}{w(\mb{r})} \Phi_i(\mb{r}) \mr{d^3}r.
\label{eq:binning}
\end{equation} 
As stated in \cite{Tegmark98}, $\Phi_i(\mb{r})$ can be localized in space -- e.g. bins or pixels -- 
or wave-like and localized in frequency space -- e.g. Fourier transforms; in any case, its application 
to data is described by Eq. \ref{eq:binning}. In this way, instead of having an infinite amount of information 
(one observed density for each point $\mb{r}$ in space), we will work with a finite set of 
bins or basis modes. This process evidently limits the information to the particular scales 
picked up by our choice of $\Phi_i(\mb{r})$, but apart from this scale choice, the final $\Psi_i(\mb{r})$ 
will not depend on the choice of $\Phi_i(\mb{r})$ \cite{Vogeley96} (assuming it forms a complete basis up to 
the chosen scale).

As we show below, we may obtain $x_i$ (possessing the properties described in Sec. \ref{sec:kl-properties}) 
from $y_i$ above through a simple linear transformation (where we use the Einstein summation convention):
\begin{equation}
x_i = \mathcal{M}_{ij}y_j.
\label{eq:xmy}
\end{equation}
Our goal then is to determine the $\mathcal{M}_{ij}$ that fulfills all three tasks in Sec. \ref{sec:kl-properties}, 
and this is achieved by describing $\mathcal{M}_{ij}$ as a product of three matrices, each one designed to 
accomplish each task without compromising previous ones. Comparing Eqs. \ref{eq:pkl-coeff}, \ref{eq:binning} and \ref{eq:xmy}, 
we see that our pKL modes $\Psi_i(\mb{r})$ will be described in terms of our choice of basis functions $\Phi_i(\mb{r})$:

\begin{equation}
\Psi_i(\mb{r}) = \mathcal{M}_{ij} \Phi_j(\mb{r}) = K^{\dagger}_{ik}W^{\dagger}_{kn}\Pi_{nj} \Phi_j(\mb{r}),
\label{eq:pkl}
\end{equation} 
that is, $\mathcal{M}_{ij}$ can be regarded as a set of coefficients for $\Phi_j(\mb{r})$ used 
to describe $\Psi_i(\mb{r})$. In the equation above, $\M{\Pi}$ makes $\Psi_i(\mb{r})$ orthogonal 
to the systematics and mean density templates $M_j(\mb{r})$, $\M{W^{\dagger}}$ pre-whitens the noise 
(makes the noise covariance matrix diagonal and uniform) and $\M{K^{\dagger}}$ diagonalizes the signal 
covariance matrix. We now describe how each of these matrices are obtained:

\begin{enumerate}

\item From Eq. \ref{eq:pkl}, we see that if
\begin{align}
\Pi_{ik} U_{kj} &= 0, \label{eq:piu} \\
U_{kj} &\equiv \int \frac{M_j(\mb{r})}{w(\mb{r})} \Phi_k(\mb{r}) \mr{d^3}r, \label{eq:u-def}
\end{align} 
then Eq. \ref{eq:m-orthogonal} is satisfied since the multiplication of 
$\M{W^{\dagger}}$ and $\M{K^{\dagger}}$ by zero still results in zero. As stated by 
\cite{Tegmark98}, there is an infinite set of matrices $\M{\Pi}$ that fulfill 
Eq. \ref{eq:piu}. Here, we stick to the simplest choice:

\begin{equation}
\M{\Pi} = \M{I} - \M{U}(\M{U}^\dagger\M{U})^{-1}\M{U}^\dagger,
\label{eq:pi-def}
\end{equation} 
where $\M{I}$ is the identity matrix. What $\M{\Pi}$ does is to project vectors describing 
the pixelized data unto a subspace orthogonal to the columns of $\M{U}$, which represent 
the pixelized systematics (and mean density). Note that $\M{\Pi}$ is Hermitian ($\M{\Pi}^\dagger = \M{\Pi}$) 
and that $\M{\Pi}^2 = \M{\Pi}$, which is a projection matrix property. Since $\M{\Pi}$ is 
a square matrix but projects vectors unto a subspace of reduced dimensionality [by $N$, 
the number of templates $M_j(\mb{r})$], it transforms linearly independent 
vectors into linear dependent ones. 

\item If we introduce Eq. \ref{eq:pkl} into Eq. \ref{eq:n-cov}, we have: 

\begin{equation}
  N_{ij} = K^{\dagger}_{il}W^{\dagger}_{lk}\Pi_{kn} N'_{nm} \Pi^\dagger_{mp}W_{pq}K_{qj} = \delta_{ij},
  \label{eq:W-derivation}
\end{equation}
\begin{equation}
 N'_{nm} \equiv \int W(\mb{r}) \frac{\bar{n}_{\mr{obs}}(\mb{r})}{w^2(\mb{r})} \Phi_n(\mb{r})\Phi_m^*(\mb{r}) \mr{d^3}r.
\label{eq:phi-noise}
\end{equation}
We can obtain our second property (uncorrelated unit noise) by choosing $\M{W}$ so

\begin{equation} 
W^{\dagger}_{ik}\Pi_{kn} N'_{nm}\Pi^\dagger_{mp}W_{pj}=\delta_{ij}  
\label{eq:whitening}
\end{equation}
and by specifying that $\M{K}$ is an unitary matrix, i.e. $K^{\dagger}_{il}K_{lj}=\delta_{ij}$, 
so its posterior application to Eq. \ref{eq:whitening} does not destroy the result. 
The process performed by $\M{W}$ is known as \emph{pre-whitening}.

Covariance matrices like $\M{N}'$ are Hermitian ($\M{A}^\dagger=\M{A}$), and thus so it is 
$\M{\Pi}\M{N}'\M{\Pi}^\dagger$. Therefore, this last matrix can be decomposed as 
$\M{\Pi}\M{N}'\M{\Pi}^\dagger = \M{Q}\M{\Lambda}\M{Q}^\dagger$, 
where $\M{Q}$ is an unitary matrix whose columns are eigenvectors (and orthogonal with respect 
to one another) of $\M{\Pi}\M{N}'\M{\Pi}^\dagger$ and $\M{\Lambda}$ is a diagonal matrix with 
the eigenvalues $\lambda_i$ of $\M{\Pi}\M{N}'\M{\Pi}^\dagger$ as diagonal elements. So we would like to 
build a matrix $\M{B}\equiv\M{Q}\M{\Lambda}^{-1/2}$, where $\M{\Lambda}^{-1/2}$ is a diagonal 
matrix with diagonal elements equal to $\lambda_i^{-1/2}$, that would transform $\M{\Pi}\M{N}'\M{\Pi}^\dagger$ 
into the identity matrix, as we desire. Unfortunately, the projection onto a subspace performed by 
$\M{\Pi}$ makes $\M{\Pi}\M{N}'\M{\Pi}^\dagger$ singular, meaning that some of its eigenvalues are zero and 
$\M{\Lambda}^{-1/2}$ cannot be computed. 

An eigenvalue of a covariance matrix corresponds to the variance of the variables when combined 
according to the associated eigenvector; and a null eigenvalue corresponds to the absence of variance that results 
from the combination of linearly dependent variables, that is, this variable combination does not carry 
any new information that was not present in previous ones. Therefore, we can eliminate this combination 
of variables from the system, which means excluding from $\M{Q}$ the columns whose eigenvalues are zero 
and from $\M{\Lambda}$ the associated rows/columns. Calling these matrices $\M{Q_0}$ and $\M{\Lambda_0}$ 
we finally have:

\begin{equation}
\M{W}=\M{Q_0}\M{\Lambda_0}^{-1/2}. 
\label{eq:w-def}
\end{equation}
For $D$ basis functions $\Phi_i(\mb{r})$ and 
$N$ systematics templates $M_j(\mb{r})$, $\M{Q_0}$ and $\M{\Lambda_0}$ have dimensions $D\times (D-N)$ and 
$(D-N)\times(D-N)$, respectively. The pre-whitening process in the pKL literature is either not 
clearly stated \cite{Tegmark98,Tegmark02,Tegmark04} or not applicable to a basis orthogonal 
to systematics \cite{Vogeley96}, and therefore we infer this to be a new approach. 

\item Lastly, we must find the matrix $\M{K}$ that diagonalizes the signal covariance matrix. 
Substituting Eq. \ref{eq:pkl} into Eq. \ref{eq:s-cov}, we get:

\begin{equation}
  S_{ij} = K^{\dagger}_{il}W^{\dagger}_{lk}\Pi_{kn} S'_{nm} \Pi^\dagger_{mp}W_{pq}K_{qj} = \lambda_{(i)}\delta_{ij},
\label{eq:s-transform}
\end{equation} 
\begin{equation}
  S'_{nm} \equiv \int W(\mb{r})W(\mb{r'}) \frac{\bar{n}_{\mr{g}}(\mb{r})}{w(\mb{r})} \frac{\bar{n}_{\mr{g}}(\mb{r'})}{w(\mb{r'})}
    \langle \delta_{\mr{g}}(\mb{r})\delta_{\mr{g}}(\mb{r'})\rangle 
    \Phi_n(\mb{r})\Phi_m^*(\mb{r'}) \mr{d^3}r\mr{d^3}r'.
\label{eq:phi-signal}    
\end{equation}
In this paper, indices inside parentheses [such as $(i)$ in Eq. \ref{eq:s-transform}] are not summed over.

As $\M{S'}$ is a covariance matrix and given that $\M{W}$ already reduced the dimensionality of 
$\M{\Pi S' \Pi^\dagger}$ by removing linearly dependent combinations of basis modes $\Phi_n(\mb{r})$, 
$\M{W^{\dagger}\Pi S' \Pi^\dagger W}$ is Hermitian and non-singular, therefore it allows the 
eigendecomposition $\M{K}\M{\Lambda}\M{K}^{\dagger}$, where the matrix we are looking for, $\M{K}$, 
is unitary, as required in the previous item.

\end{enumerate} 

Given $\M{\Pi}$, $\M{W}$ and $\M{K}$ we can build $\Psi_i(\mb{r})$ (Eq. \ref{eq:pkl}) whose coefficients 
$x_i$ (Eq. \ref{eq:pkl-coeff}) have diagonal signal and noise covariance matrices. Since the noise 
is uniform, modes with highest signal variance $\lambda_{(i)}$ are modes with the highest signal-to-noise 
ratio (SNR). We can therefore select a fraction of the calculated modes and extract the most information out 
of a small set of modes $\Psi_i(\mb{r})$.  

\section{The ``pseudo'' in pKL: building a general angular part first} 
\label{sec:pseudo-kl}

The program presented in Sec. \ref{sec:kl-method} is complete from the theoretical point of view, 
but it may suffer from an implementation problem: the computation of potentially $D\times D$ 
~$\mathcal{M}_{ij}$ terms that describe the $\Psi_i(\mb{r})$ modes in terms of $\Phi_j(\mb{r})$ 
(Eq. \ref{eq:pkl}) requires eigenvector decompositions, a task whose computing time and storage 
increases as $D^3$ and $D^2$, respectively. If each spatial dimension were described by 50 independent modes, 
we would have $D=125000$. One proposal to ease implementation is to assume that 
$\Psi_i(\mb{r})=\Psi_i^{\mr{z}}(r)\Psi_i^{\mr{\theta}}(\mb{\hat{r}})$ is separable \cite{Tegmark02} into radial 
$\Psi_i^{\mathrm{z}}(r)$ and angular $\Psi_i^{\mathrm{\theta}}(\mb{\hat{r}})$ parts [implying the same for the 
basis functions $\Phi_j(\mb{r})$], and to do similarly to 
the selection function $\bar{n}_{\mr{g}}(\mb{r})=\bar{n}_{\mr{g}}^{\mr{z}}(r)\bar{n}_{\mr{g}}^{\mr{\theta}}(\mb{\hat{r}})$, 
the systematics templates $M_j(\mb{r})=M^{\mr{z}}_j(r)M^{\mr{\theta}}_j(\mb{\hat{r}})$ and the 
weights $w(\mb{r})=w_{\mr{z}}(r)w_{\mr{\theta}}(\mb{\hat{r}})$. 
More importantly, the proposal is to calculate $\Psi_i^{\mr{\theta}}(\mb{\hat{r}})$ in 
an independent way from $\Psi_i^{\mr{z}}(r)$. Once we have $\Psi_i^{\mr{\theta}}(\mb{\hat{r}})$ in hand 
and have selected only those with the highest SNR, we proceed to compute a set of radial modes 
$\Psi_i^{\mr{z}}(r)$ for each specific angular mode. The fact that we have previously selected a fraction of 
angular modes before computing the radial ones decreases the amount of computing time. Moreover, 
the computation of radial modes can be done in parallel.

Under this approach, we can completely separate Eq. \ref{eq:m-orthogonal}:

\begin{equation}
\int \frac{M^{\mr{z}}_j(r)}{w_{\mr{z}}(r)} \Psi^{\mr{z}}_k(r) r^2 \mr{d}r
\int \frac{M^{\mr{\theta}}_j(\mb{\hat{r}})}{w_{\mr{\theta}}(\mb{\hat{r}})} \Psi^{\mr{\theta}}_k(\mb{\hat{r}}) \mr{d^2}\hat{r} = 0,
\label{eq:m-orthogonal-sep}
\end{equation}
and we see that it is enough having either the angular or the radial modes orthogonal to their mean density and 
systematics templates (although we will require both just to be on the safe side), and these orthogonalizations may be achieved 
by radial and angular projection matrices $\M{\Pi}_{\mr{z}}$ and $\M{\Pi}_{\mr{\theta}}$ that combine radial $\Phi^{\mr{z}}_k(r)$ or 
angular $\Phi_i^{\mr{\theta}}(\mb{\hat{r}})$ basis modes. 

If we assume that $\bar{n}_{\mr{obs}}(\mb{r})$ is also separable [which will not be the case if, 
for instance, $\bar{n}_{\mr{g}}(\mb{r})$ and $s(\mb{r})$ do not have either the same radial or angular parts], 
then we can also completely separate \ref{eq:phi-noise}:

\begin{equation}
  N'_{[in][jm]} = 
  \int W_{\mr{z}}(r) \frac{\bar{n}^{\mr{z}}_{\mr{obs}}(r)}{w^2_{\mr{z}}(r)} 
  \Phi^{\mr{z}}_i(r)\Phi^{\mr{z}*}_j(r) r^2\mr{d}r
  \int W_{\mr{\theta}}(\mb{\hat{r}}) \frac{\bar{n}^{\mr{\theta}}_{\mr{obs}}(\mb{\hat{r}})}{w^2_{\mr{\theta}}(\mb{\hat{r}})} 
  \Phi_n^{\mr{\theta}}(\mb{\hat{r}})\Phi_m^{\mr{\theta}*}(\mb{\hat{r}}) \mr{d^2}\hat{r}.
  \label{eq:phi-noise-sep}
\end{equation} 
On the left side, we wrote the radial and angular indices inside square brackets to emphasize that, together, 
they represent a single dimension in $\M{N'}$ (i.e. $\M{N'}$ will be regarded as a block matrix). To 
pre-whiten $\M{\Pi N' \Pi^{\dagger}}$ and transform it into an identity matrix, we can pre-whiten the angular and radial 
parts separately.

Unfortunately, the separation obtained in Eqs. \ref{eq:m-orthogonal-sep} and \ref{eq:phi-noise-sep} 
is not feasible for $\M{S'}$ due to $\langle \delta_{\mr{g}}(\mb{r})\delta_{\mr{g}}(\mb{r'})\rangle $. 
In other words, the fact that the galaxy correlation function 
$\xi_{\mr{g}}(|\mb{r}-\mb{r'}|) = \langle \delta_{\mr{g}}(\mb{r})\delta_{\mr{g}}(\mb{r'})\rangle $ cannot be described 
as a product of angular and radial parts precludes the separability of the signal covariance matrix $\M{S'}$.
One solution is to compute optimal angular modes 
$\Psi_i^{\mr{\theta}}(\mb{\hat{r}})$ for the \emph{projected} density, hoping the result will approximate 
the optimal angular functions for the unprojected density. The computation of such angular modes is analogue 
to setting the radial part of $\Phi_n(\mb{r})$ to unity in Eq. \ref{eq:phi-signal} and then solving Eq. 
\ref{eq:s-transform}. Note that, since Eq. \ref{eq:phi-noise-sep} is separable, the $\M{N'}$ matrix for 
the angular part (last integral in Eq. \ref{eq:phi-noise-sep}) is the same as the $\M{N'}$ matrix for 
the projected density, apart from a overall constant [which does not affect the outcome of $\Psi_i^{\mr{\theta}}(\mb{\hat{r}})$].

\subsection{Optimal angular modes}
\label{sec:pkl-ang}

It is clear that the choice of $w_{\mr{z}}(r)$ used when projecting the number density of galaxies
changes the resulting angular modes since it will emphasize the clustering at certain radial distances 
over others. One question we tackle in this paper is: what are the optimal weights 
for obtaining the angular modes?
 
To answer this question, we proposed two educated guesses: 

\begin{enumerate}[i)]

\item That extracting the maximum amount of information from the correlation of a 3D field is 
equivalent to extracting the maximum amount of information from all correlations between tomographic 
slices of this field; this maximization for each pair of slices can be achieved by KL modes built 
specifically for them, using the same methodology described in Sec. \ref{sec:kl-method}. In this 
process, we can make the pre-whitening matrix $\M{W}$ to be the same for every pair of slices if we adopt 
the radial weights $w_{\mr{z}}(r)=r\sqrt{\bar{n}^{\mr{z}}_{\mr{obs}}(r)}$. In this way, the only difference 
between KL modes for different pairs would come from the matrices $\M{S'}$.

\item Given the impossibility that a single set of angular modes can extract the maximum amount 
of information for all slice combinations (since $\M{S'}$ would be different for each one), the 
one that gets closest to this task would be the one extracted from the average of the $\M{S'}$ matrices; 
this corresponds to angular modes obtained from the projected density weighted by:
$w_{\mr{z}}(r)=r\sqrt{\bar{n}^{\mr{z}}_{\mr{obs}}(r)}$.

\end{enumerate}
As our basis angular functions we will choose spherical harmonics:
\begin{equation} 
\Phi_i^{\mr{\theta}}(\mb{\hat{r}})=Y^*_{\ell_i m_i}(\mb{\hat{r}}).
\end{equation}
Under these choices, the matrix $\M{\Pi_{\mr{\theta}}}$ used to build the angular modes is obtained 
from Eq. \ref{eq:pi-def}, where the elements of $\M{U}$ are given by:

\begin{equation}
U^{\mr{\theta}}_{kj} = \int \frac{M^{\mr{\theta}}_j(\mb{\hat{r}})}{w_{\mr{\theta}}(\mb{\hat{r}})}Y^*_{\ell_k m_k}(\mb{\hat{r}})\mr{d^2}\hat{r}.
\end{equation}
We see that the columns of $\M{U_{\mr{\theta}}}$ are the spherical harmonic coefficients of 
$M^{\mr{\theta}}_j(\mb{\hat{r}})/w_{\mr{\theta}}(\mb{\hat{r}})$.
  
The pre-whitening matrix $\M{W}$ is obtained from the process described in Sec. \ref{sec:kl-mechanics} (Eq. \ref{eq:w-def}), 
but starting from:

\begin{equation}
  N'_{ij} = 
  \int W_{\mr{\theta}}(\mb{\hat{r}}) \frac{\bar{n}^{\mr{\theta}}_{\mr{obs}}(\mb{\hat{r}})}{w^2_{\mr{\theta}}(\mb{\hat{r}})} 
  Y^*_{\ell_i m_i}(\mb{\hat{r}})Y_{\ell_j m_j}(\mb{\hat{r}}) \mr{d^2}\hat{r}.
\end{equation}
Here we will define the functional $J_{\ell'm'\ell m}$:
\begin{equation}
  \begin{split}
    & J_{\ell'm'\ell m}[f] \equiv \int f(\mb{\hat{r}}) Y_{\ell' m'}^*(\mb{\hat{r}}) Y_{\ell m}(\mb{\hat{r}}) \mr{d^2}\hat{r} 
    = \sum_{LM}f_{LM} \int Y_{LM}(\mb{\hat{r}}) Y_{\ell m}(\mb{\hat{r}}) Y_{\ell' m'}^*(\mb{\hat{r}}) \mr{d^2}\hat{r} \\
    & = \sum_{LM}f_{LM} \sqrt{\frac{(2L+1)(2\ell+1)(2\ell'+1)}{4\pi}}(-1)^{m'} 
    \begin{pmatrix}
      L & \ell & \ell' \\
      0 & 0    & 0  
    \end{pmatrix}
    \begin{pmatrix}
      L & \ell & \ell' \\
      M & m    & -m'   
    \end{pmatrix},
  \end{split}
\label{eq:def-Jlmlm}
\end{equation}
where $f_{LM}\equiv\int f(\mb{\hat{r}})Y^*_{LM}\mr{d^2}\hat{r}$ and 
$\begin{pmatrix} L & \ell & \ell' \\ M & m & m'\end{pmatrix}$ are the Wigner 3-$j$ symbols. Thus, 
we have:
\begin{equation}
  N'_{ij} = J_{\ell_i m_i \ell_j m_j}\left[\frac{W_{\mr{\theta}}\bar{n}^{\mr{\theta}}_{\mr{obs}}}{w^2_{\mr{\theta}}}\right].
\end{equation}
We also point out that, according to its definition (Eq. \ref{eq:def-Jlmlm}), $J_{\ell'm'\ell m}$ is Hermitian,
that is: $J_{\ell'm'\ell m} = J^*_{\ell m \ell' m'}$.

Finally, the angular modes' matrix $\M{K}$ is computed following Sec \ref{sec:kl-mechanics}, 
using an $\M{S'}$ matrix with elements:
\begin{equation}
  S'_{ij} = \int W_{\mr{\theta}}(\mb{\hat{r}}) W_{\mr{\theta}}(\mb{\hat{r'}})
  \frac{\bar{n}^{\mr{\theta}}_{\mr{g}}(\mb{\hat{r}})}{w_{\mr{\theta}}(\mb{\hat{r}})} 
  \frac{\bar{n}^{\mr{\theta}}_{\mr{g}}(\mb{\hat{r}'})}{w_{\mr{\theta}}(\mb{\hat{r}'})}
    \langle \sigma_{\mr{g}}(\mb{\hat{r}})\sigma_{\mr{g}}(\mb{\hat{r}'})\rangle 
    Y^*_{\ell_i m_i}(\mb{\hat{r}})Y_{\ell_j m_j}(\mb{\hat{r}'}) 
    \mr{d^2}\hat{r}\mr{d^2}\hat{r}',
\label{eq:ang-signal-1}
\end{equation}
where $\sigma_{\mr{g}}(\mb{\hat{r}})$ is the density projected under radial weights $w_{\mr{z}}(r)=r\sqrt{\bar{n}^{\mr{z}}_{\mr{obs}}(r)}$:
\begin{equation}
\sigma_{\mr{g}}(\mb{\hat{r}}) \equiv \int W_{\mr{z}}(r)\frac{\bar{n}^{\mr{z}}_{\mr{g}}(r)}{w_{\mr{z}}(r)}
\delta_{\mr{g}}(\mb{r})r^2\mr{d}r.
\label{eq:proj-density}
\end{equation}
Eq. \ref{eq:ang-signal-1} can be written in terms of the angular power spectrum 
$C_{[\ell m][\ell'm']}\equiv C_{(\ell)}\delta_{\ell\ell'}\delta_{mm'}$ of the full-sky 
projected density $\sigma_{\mr{g}}(\mb{\hat{r}})$:
\begin{equation}
S'_{ij} = J_{[\ell_i m_i][\ell m]}\left[\frac{W_{\mr{\theta}}\bar{n}^{\mr{\theta}}_{\mr{g}}}{w_{\mr{\theta}}}\right]
C_{[\ell m][\ell' m']}
J^\dagger_{[\ell' m'][\ell_j m_j]}\left[\frac{W_{\mr{\theta}}\bar{n}^{\mr{\theta}}_{\mr{g}}}{w_{\mr{\theta}}}\right].
\label{eq:phi-ang-signal}
\end{equation}
In the equation above, remember that indices inside square brackets actually describe a single dimension 
of the matrix and repeated indices get summed over (Eq. \ref{eq:phi-ang-signal} is a matrix multiplication).

We thus see that the impact of the radial weights $w_{\mr{z}}(r)$ on the angular modes boils down to 
changing the full-sky angular power spectrum in Eq. \ref{eq:phi-ang-signal}. It is interesting to note 
that $C_\ell$ refers to an isotropic property and does not depend on the shape, size or orientation of 
the survey's mask (or other angular properties). Therefore, different choices of a fiducial $C_\ell$,
used to build the KL modes, conserve all their orthogonality properties: an optimal choice only leads 
to higher SNR for the selected modes.

\begin{figure}
  \center
  \includegraphics[width=0.49\textwidth]{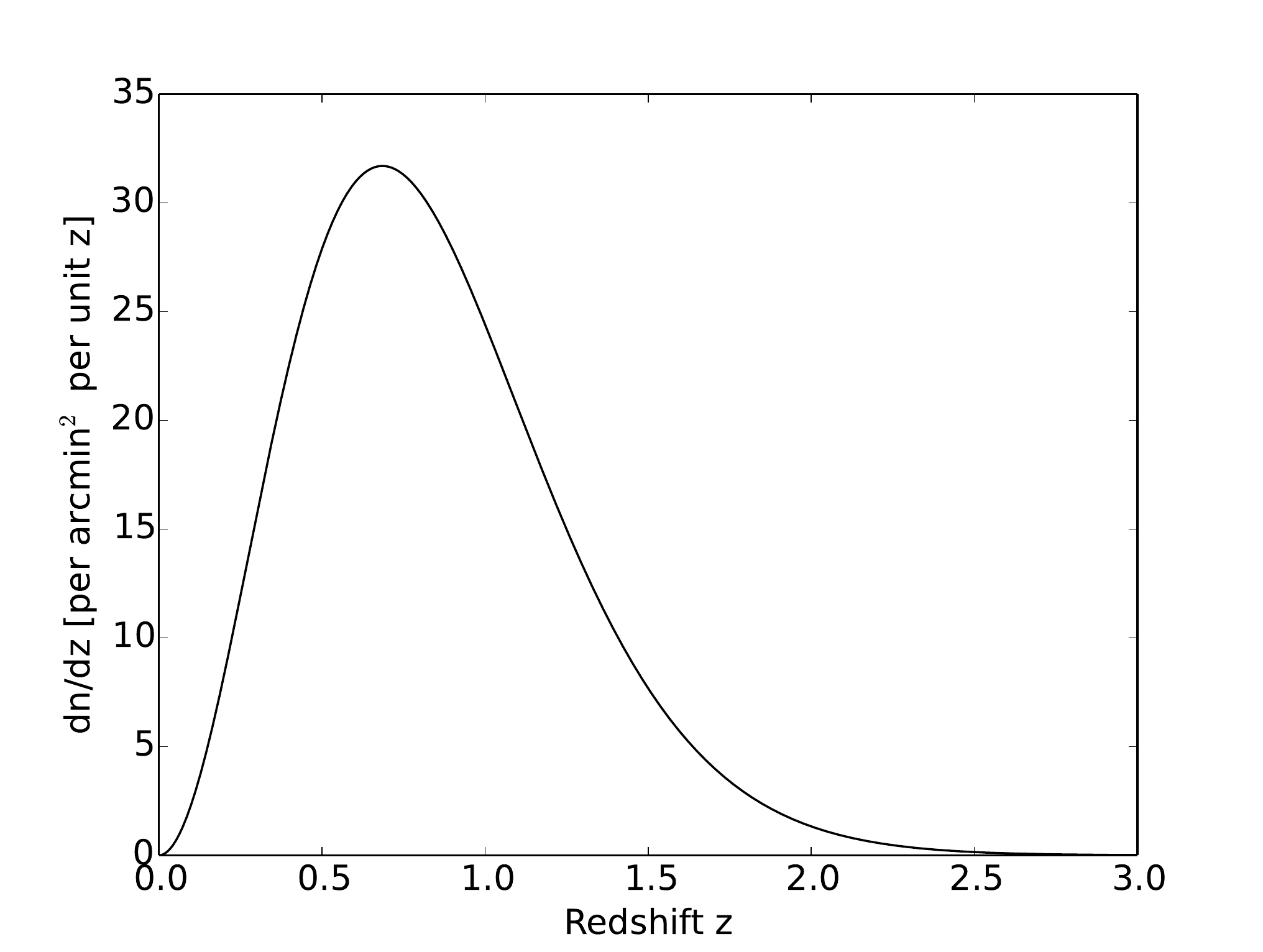}
  \includegraphics[width=0.49\textwidth]{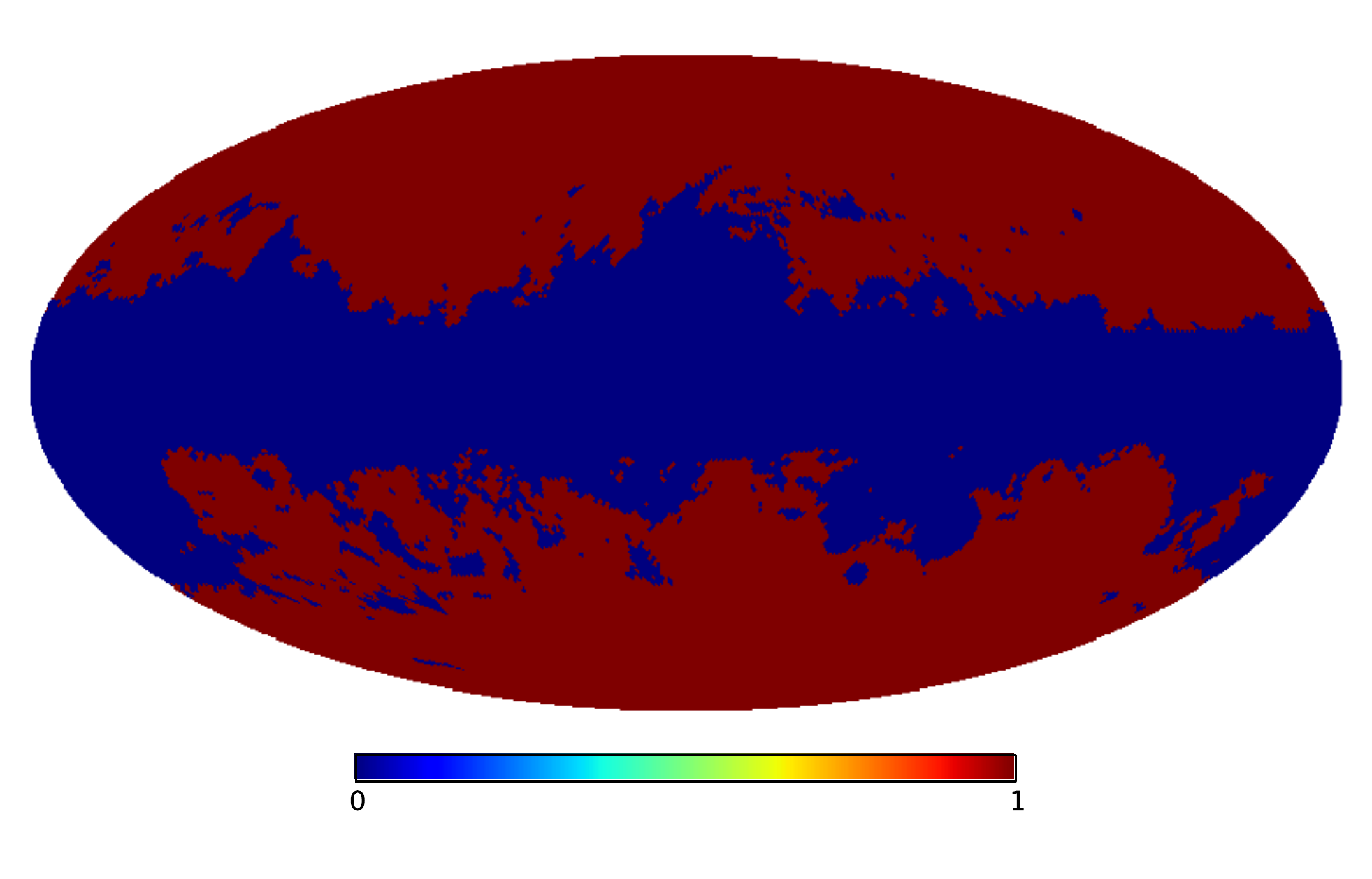}
  \caption{Selection function used in our numerical tests of different radial weightings. 
      The left panel shows radial part (similar to the one expected from the Euclid survey), 
      while the right panel shows the angular part, that removes regions 
      of high stellar density and extinction, in Galactic coordinates. The angular mask was 
      based in the one used in \cite{Bengaly18,Novaes18}.} 
  \label{fig:fiducial-selection}
\end{figure}

To verify that our choice of $w_{\mr{z}}(r)$ -- and, consequently, the weighted projected $C_\ell$ -- 
has a positive impact on the extracted SNR, we compared the variances of the signal (computed numerically 
while the noise was set to unity) 
inside 16 thin slices in the redshift range $0.07<z<1.82$ for two different sets of angular KL modes: 
one built from our radial weighting and another from a simple projection of all galaxies into a single map 
[i.e. $w_{\mr{z}}(r)=1$]. As the radial and angular selection functions $\bar{n}^{\mr{z}}_{\mr{obs}}(r)$ and 
$W_{\mr{\theta}}(\mb{\hat{r}})\bar{n}^{\mr{\theta}}_{\mr{obs}}(\mb{\hat{r}})$, we adopted the Euclid-like 
redshift distribution and the almost full-sky binary mask shown in Fig. \ref{fig:fiducial-selection}.

\begin{figure}
  \center
  \includegraphics[width=0.7\textwidth]{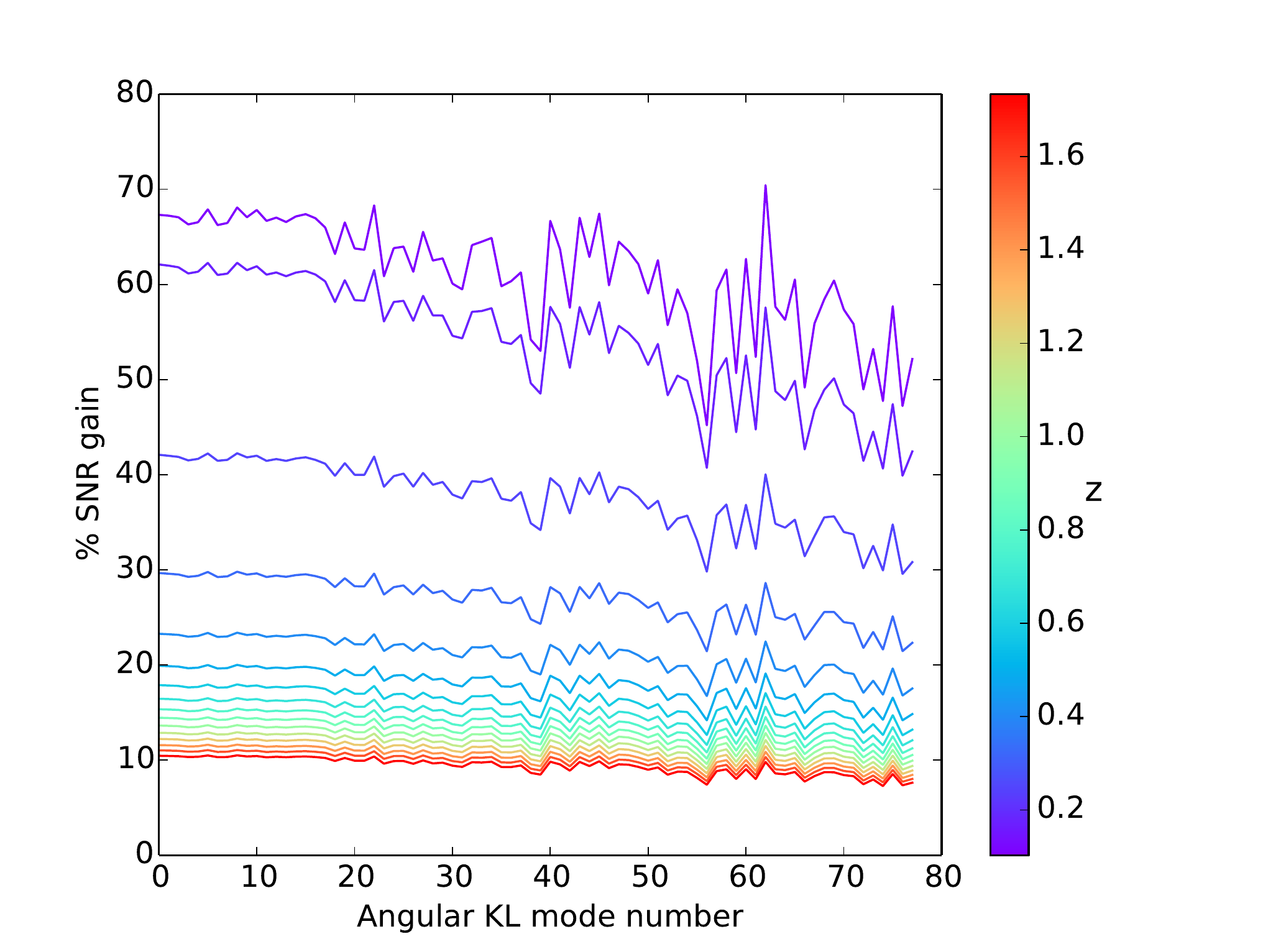}
  \caption{Fractional difference between the SNRs $S_{\mr{w}}$ and $S_{\mr{u}}$ ($S_{\mr{w}}/S_{\mr{u}}-1$) 
      extracted from 16 thin redshift slices using angular KL modes built with weighted and unweighted 
      projected $C_\ell$s, as a function of the first 78 KL modes (ordered by signal variance). The SNRs were computed 
      with Eqs. \ref{eq:s-transform} and \ref{eq:phi-ang-signal} for each bin (colored according to 
      their central redshift value) using the bin's auto-$C_\ell$s.} 
  \label{fig:snr-gain}
\end{figure}

Fig. \ref{fig:snr-gain} shows that there is a significant gain in the SNR for all modes and 
redshifts when the angular KL modes are built from the $C_\ell$ of the projected density that 
uses our radial weights. The reason is that this $C_\ell$ has a shape closer to the thin slices' 
$C_\ell$s than the unweighted one. We point out that the $C_\ell$ overall amplitude does not alter the 
derived KL modes. 

It is worth pointing out that, in most cosmological models (as well as in the standard model), 
scales with the largest signal are intermediate ones. Consequently, these will be the scales probed by the 
KL modes constructed from a $C_\ell$ based on actual expected data. In case we want to build modes to probe the 
largest scales, one solution is to adopt a fiducial $C_\ell$ with boosted power on these scales. As mentioned 
above, this will not affect the orthogonality of the modes inside the survey's mask and with respect to systematics' 
maps, although it will turn the optimal radial weighting innocuous. Lastly, the RSD and lensing effects tend to 
increase the power on the largest scales \cite{Challinor11}, so the boosted fiducial $C_\ell$ is likely close 
to optimal.

The results from this section does not only serve as the angular part of 3D pKL modes, but also as a full KL 
method to analyze 2D, projected data. Moreover, our discussion of optimal weighting also applies to the quest 
of building a single set of angular KL modes to extract information for a set of tomographic slices of 3D cosmological 
fields.

\subsection{Light-cone effects and new radial modes}
\label{sec:pkl-radial}

Once the angular pKL modes $\Psi_i^{\mr{\theta}}(\mb{\hat{r}})$ are specified, we can write 

\begin{equation}
\Psi_{[ij]}(\mb{r})=\mathcal{M}^{(i)}_{jn}\Phi_n^{\mr{z}}(r)\Psi_{(i)}^{\mr{\theta}}(\mb{\hat{r}}) = 
\mathcal{M}^{\mr{\theta}}_{ip}\mathcal{M}^{(i)}_{jn} \Phi_n^{\mr{z}}(r)\Phi_{p}^{\mr{\theta}}(\mb{\hat{r}}),
\label{eq:psi-radial}
\end{equation}
remembering that we adopt the Einstein summation convention except for indices inside parentheses. 
In Eq. \ref{eq:psi-radial} we also chose to name the final pKL mode using a compound index made of two auxiliary 
indices, one for the angular part and the other for the radial part. What Eq. \ref{eq:psi-radial} says is 
that, for each angular mode $\Psi_i^{\mr{\theta}}(\mb{\hat{r}})$ held fixed, we will build a set of 3D modes 
$\Psi_{[ij]}(\mb{r})$ (where $i$ and $j$ select the angular and radial part, respectively) whose radial 
part is a linear combination of radial basis functions $\Phi_n^{\mr{z}}(r)$. Note that this linear combination 
is different for each angular mode, even if the radial index $j$ is the same, e.g.: 
$\mathcal{M}^{(1)}_{jn} \neq \mathcal{M}^{(2)}_{jn}$.

The choice for the radial basis functions $\Phi_n^{\mr{z}}(r)$ made in \cite{Tegmark02} 
was logarithmic waves \cite{Hamilton96}: 

\begin{equation}
Z_\omega(r) \equiv \frac{1}{\sqrt{2\pi}}\frac{e^{-i\omega\ln(r)}}{r^{3/2}}. 
\end{equation}
This choice simplifies further calculations \emph{if} light-cone effects can be ignored (i.e. if all galaxies 
in the survey are observed at the same Universe's age) and if $\alpha(r)$, defined below, can be considered constant:
\begin{equation}
\alpha(r) \equiv \frac{\partial \ln [r^2 \bar{n}_{\mr{g}}(r)]}{\partial \ln r}. 
\end{equation}
On the downside, $Z_\omega(r)$ are only orthogonal inside the interval $[0,\infty)$ and $\omega$, 
the number that characterizes the mode's scale, is real and continuous. This last property is 
particularly unpleasant since it does not clear out which modes -- and how many -- are required to 
describe $\Psi^{\mr{z}}_i(r)$ up to a certain scale of interest. For this reason -- and given that we 
take into consideration light-cone effects -- we suggest the use of 
radial basis functions $\Phi_n^{\mr{z}}(r)$ that completely describes functions in a finite interval. 
Possible choices are discrete Fourier series, Legendre or Chebyshev polynomials 
or top-hat bins.

Legendre and Chebyshev polynomials have the advantage they are defined in a (non-periodic) finite interval 
(in contrast to Fourier series, which are periodical). Therefore, they may provide better 
descriptions for functions that would be discontinuous at periodic boundaries [e.g. $\bar{n}_{\mr{g}}(r)$], 
as they do not suffer from the Gibbs phenomenon. On the other hand, discrete Fourier series or top-hat functions 
might be easier to integrate (due to the use of fast Fourier transforms in the first and 
to the avoidance of oscillatory functions in the last case). Despite this choice, the final radial KL modes 
should be independent of the basis used (up to a certain scale). 

To compute the radial $\mathcal{M}^{(i)}_{jn}$, we follow the usual procedure, described in Sec. 
\ref{sec:kl-mechanics}, observing that it is not required to adopt the same choice for 
$w_{\mr{z}}(r)$ as in Sec. \ref{sec:pkl-ang}, now that $\Psi_{(i)}^{\mr{\theta}}(\mb{\hat{r}})$ is 
already built. Independently from the chosen radial basis, the first step is the orthogonalization 
with respect to the radial component of the systematics, $M^{\mr{z}}_j(r)$. As usual, we will use 
Eq. \ref{eq:pi-def} with elements of $\M{U}$ given by: 
\begin{equation}
U^{\mr{z}}_{kj} = \int \frac{M^{\mr{z}}_j(r)}{w_{z}(r)}\Phi_k^{\mr{z}}(r)r^2\mr{d}r.
\end{equation}

For separable $\bar{n}_{\mr{obs}}(\mb{r})$, the pre-whitening matrix $\M{W}$ is obtained from the 
procedure described in Sec. \ref{sec:kl-mechanics}, starting from:

\begin{equation}
  N'_{ij} = \int W_{\mr{z}}(r) \frac{\bar{n}^{\mr{z}}_{\mr{obs}}(r)}{w^2_{\mr{z}}(r)} \Phi_i^{\mr{z}}(r)\Phi_j^{\mr{z}*}(r) r^2\mr{d}r.
\end{equation}
In this case, both $\M{\Pi}$ and $\M{W}$ are the same for every angular mode 
$\Psi_{i}^{\mr{\theta}}(\mb{\hat{r}})$, since $\M{U}$ and $\M{N'}$ are separable.

Since the signal covariance matrix is non-separable, each angular mode will require its own $\M{K}$ matrix. 
To compute them, we assume that the preparation of optimal angular pKL modes (Sec. \ref{sec:pkl-ang}) 
already made the covariance signal matrix for basis modes $\Phi_n^{\mr{z}}(r)\Psi_{i}^{\mr{\theta}}(\mb{\hat{r}})$ 
sufficiently close to zero for components whose angular parts are different 
[e.g. $\Phi_n^{\mr{z}}(r)\Psi_{1}^{\mr{\theta}}(\mb{\hat{r}})$ and $\Phi_m^{\mr{z}}(r)\Psi_{2}^{\mr{\theta}}(\mb{\hat{r}})$ 
for any $m,n$].
Therefore, we only need to diagonalize the matrix $\M{W^{\dagger}\Pi S' \Pi^\dagger W}$ for the same 
$\Psi_i^{\mr{\theta}}(\mb{\hat{r}})$. We start by computing $\M{S'}$.

We will begin by relating the galaxy density contrast $\delta_{\mr{g}}(\mb{r})$ directly accessible by 
observations -- i.e. in redshift space and on the light-cone -- to the matter density contrast in configuration 
space and at a fixed time, $\delta_0(\mb{r})$. Assuming a linear galaxy bias $b(r)$ (that depends on 
$r$ both due to luminosity bias \cite{Percival04} and due to galaxy evolution) and linear perturbation 
theory, this relation can be written as $\delta_{\mr{g}}(\mb{r}) = \hat{S} \delta_0(\mb{r})$, where 
the RSD operator $\hat{S}$ is given by \citep{Matsubara00}:

\begin{equation}
\hat{S} = b(r)D(r)+\left[ \gamma(r)\frac{\partial}{\partial r} + D(r)f(r) \frac{\partial^2}{\partial r^2}\right] \nabla^{-2}.
\label{eq:rsd-operator}
\end{equation}
In the equation above, $D(r)$ is the matter growth function (described in terms of $r$ since the observations are made 
on the light-cone), $f(r)$ is its logarithmic derivative $f\equiv\mathrm{d}\ln D/\mathrm{d}\ln a$ in terms of the scale 
factor $a$, $\nabla^{-2}$ is the inverse of the Laplacian operator [easier to implement when $\delta_0(\mathbf{r})$ is 
described in Fourier space], and $\gamma(r)$ is:

\begin{equation}
\gamma(r) \equiv \frac{2D(r)f(r)}{r} + 
\frac{1}{\bar{n}^{\mr{z}}_{\mr{g}}(r)}\frac{\partial[D(r)f(r)\bar{n}^{\mr{z}}_{\mr{g}}(r)]}{\partial r}.
\end{equation}

The second step is to express $\delta_0(\mathbf{r})$ by its Fourier transform $\tilde{\delta}_0(\mathbf{k})$, such that:

\begin{equation}
\nabla^{-2}\delta_0(\mathbf{r}) = 
\frac{1}{(2\pi)^3}\int\frac{-\tilde{\delta}_0(\mathbf{k})}{k^2} e^{i\mb{k}\cdot \mb{r}}\mr{d^3}k;
\label{eq:inv-laplacian}
\end{equation}
and the third step is to expand $e^{i\mb{k}\cdot \mb{r}}$ in spherical waves 
[i.e. spherical Bessel functions $j_\ell(x)$ and spherical harmonics]:

\begin{equation}
e^{i\mb{k}\cdot \mb{r}} = 4\pi \sum_{\ell=0}^\infty\sum_{m=-\ell}^\ell i^\ell j_\ell(kr) Y^*_{\ell m}(\mb{\hat{k}}) Y_{\ell m}(\mb{\hat{r}}).
\label{eq:ekr-expansion}
\end{equation}
With Eqs. \ref{eq:rsd-operator}, \ref{eq:inv-laplacian} and \ref{eq:ekr-expansion}, we can compute 
$\delta_{\mr{g}}(\mb{r}) = \hat{S} \delta_0(\mb{r})$:

\begin{equation}
\delta_{\mr{g}}(\mb{r}) = \frac{1}{2\pi^2} \sum_{\ell,m} i^\ell 
\int \tilde{\delta}_0(\mathbf{k}) G_\ell(k,r) Y^*_{\ell m}(\mb{\hat{k}}) Y_{\ell m}(\mb{\hat{r}}) \mr{d^3}k,
\label{eq:delta-g}
\end{equation}

\begin{equation}
G_\ell(k,r) \equiv b(r)D(r)j_\ell(kr) - \gamma(r)\frac{j'_\ell(kr)}{k}-D(r)f(r)j''_{\ell}(kr),
\label{eq:Glkr-def}
\end{equation}
where $j'_\ell(x)$ and $j''_\ell(x)$ are $j_\ell(x)$'s first and second derivatives. 

Let us call $s_{[ij]}$ (for `signal') the $x_{[ij]}$ term proportional to $\delta_{\mr{g}}(\mb{r})$:

\begin{equation}
s_{[ij]} \equiv \int W(\mb{r}) \frac{\bar{n}_{\mr{g}}(\mb{r})}{w(\mb{r})} \delta_{\mr{g}}(\mb{r}) \Psi_{[ij]}(\mb{r}) \mr{d^3}r.
\label{eq:s-def}
\end{equation}

By inserting Eqs. \ref{eq:psi-radial} and \ref{eq:delta-g} into Eq. \ref{eq:s-def} and inverting the order of 
the integrals on $\mb{r}$ and $\mb{k}$, we get:

\begin{equation}
s_{[ij]} = \frac{\mathcal{M}^{(i)}_{jn}}{2\pi^2} \sum_{\ell,m} i^\ell 
\int \tilde{\delta}_0(\mb{k}) \tilde{G}_{\ell m}^{in}(k) Y^*_{\ell m}(\mb{\hat{k}}) \mr{d^3}k,
\label{eq:s-final}
\end{equation}

\begin{equation}
\tilde{G}_{\ell m}^{in}(k) \equiv \mathcal{M}^\theta_{ip} 
J_{[\ell_p m_p][(\ell m)]}\left[\frac{W_{\mr{\theta}}\bar{n}^{\mr{\theta}}_{\mr{g}}}{w_{\mr{\theta}}}\right] 
\int W_{\mr{z}}(r) \frac{\bar{n}_{\mr{g}}^{\mr{z}}(r)}{w_{\mr{z}}(r)} \Phi^{\mr{z}}_n(r) G_\ell(k,r) r^2\mr{d}r.
\end{equation}
Remembering that the matter power spectrum at a fixed time $P_0(k)$ is defined by:
\begin{equation}
\langle\tilde{\delta}_0(\mathbf{k})\tilde{\delta}^*_0(\mathbf{k'})\rangle = 
(2\pi)^3P_0(k)\delta_{\mr{D}}^3(\mb{k}-\mb{k'}),
\end{equation}
we can compute the signal covariance matrix $S_{[ij][np]}$ of the 3D pKL coefficients:

\begin{equation}
S_{[ij][np]} = \langle s_{[ij]}s^*_{[np]} \rangle = \mathcal{M}^{(i)}_{jq}~ 
\frac{2}{\pi} \int P_0(k)\sum_{\ell, m}\tilde{G}_{\ell m}^{iq}(k) \tilde{G}_{\ell m}^{nr*}(k)k^2\mr{d}k 
~\mathcal{M}^{(n)\dagger}_{rp}.
\label{eq:pkl-signal-cov}
\end{equation}
In the above equation, the term on the right is of the kind $\mathcal{M}\M{S'}\mathcal{M^\dagger}$,
where $\mathcal{M}$ is the matrix of coefficients that describes radial pKL modes in terms of 
the radial basis functions, given a certain angular pKL mode (set by the indices $i$ and $n$). 
If we want to solve Eq. \ref{eq:pkl-signal-cov} for $\mathcal{M}^{(i)}_{jq}$, the problem is 
overdetermined, since this same set of coefficients must solve for all $n$ in $\mathcal{M}^{(n)\dagger}_{rp}$
(when $i\neq n$ we expect $S_{[ij][np]} = 0$ since these are off-diagonal terms). However, we assume 
that the diagonalization of the angular modes' signal covariance matrix already made these terms 
sufficiently close to zero, so we can find the radial coefficients $\mathcal{M}^{(i)}_{jq}$ by 
setting $n=i$ in Eq. \ref{eq:pkl-signal-cov} and following the standard procedure 
(third item of Sec. \ref{sec:kl-mechanics}). 

\begin{figure}
  \center
  \includegraphics[width=1.0\textwidth]{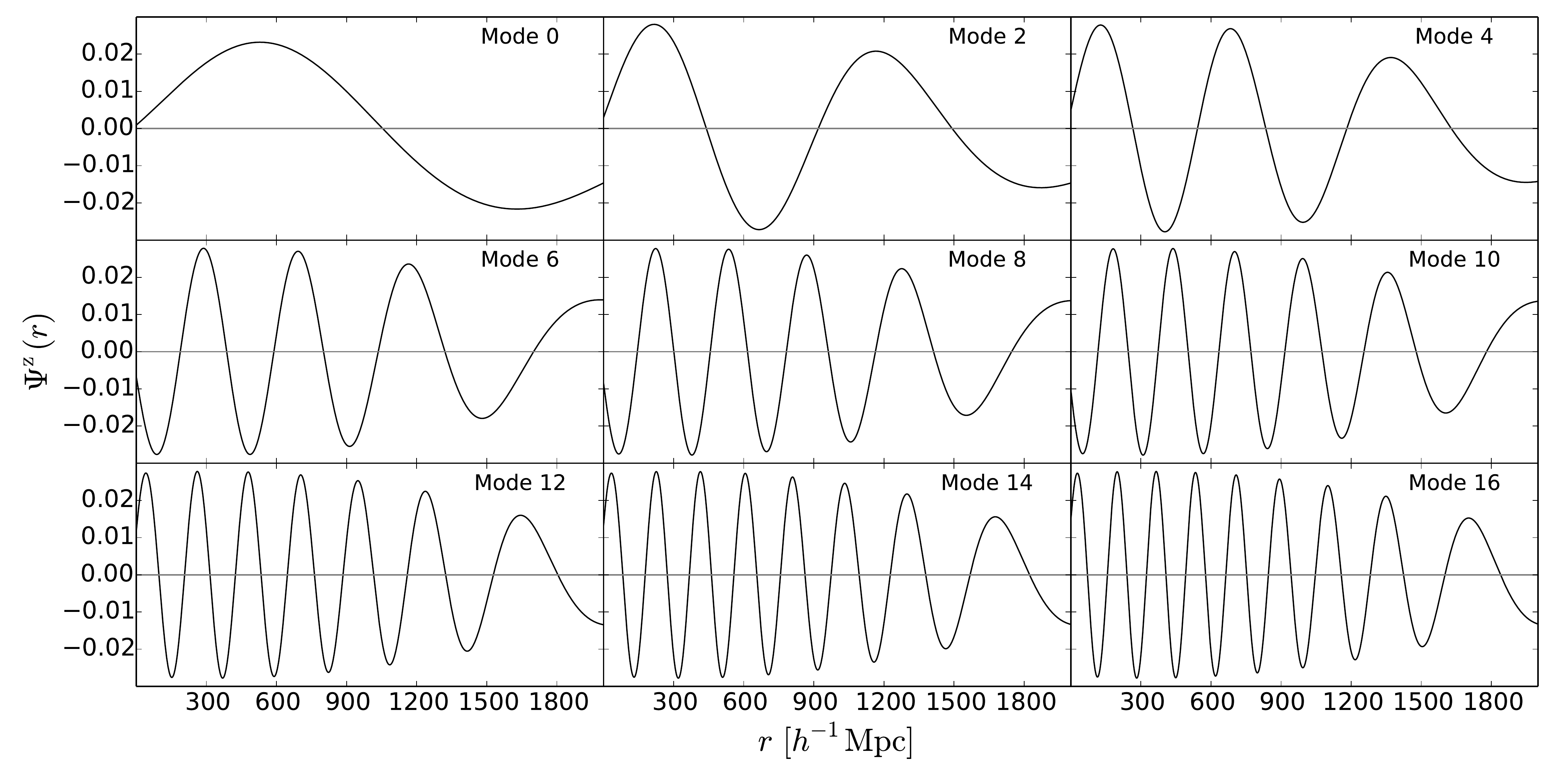}
  \caption{A few radial KL modes $\Psi_i^{\mr{z}}(r)$ computed for $\Psi_0^{\mr{\theta}}(\mb{\hat{r}})=1$. 
      We adopted the radial weights $w_{\mr{z}}(r)=r\sqrt{\bar{n}^{\mr{z}}_{\mr{obs}}(r)}$, the redshift 
      distribution shown in Fig. \ref{fig:fiducial-selection}, 2000 top-hat basis functions covering the 
      redshift range $0<z<2$, $P_0(k)=4000 (k/0.1h\mr{Mpc^{-1}})^{-2.5}$ and remaining $\Lambda$CDM 
      cosmological parameters given by Planck \cite{Planck16}.} 
  \label{fig:radial-kl-modes}
\end{figure}

Fig. \ref{fig:radial-kl-modes} shows examples of radial KL modes, computed for the largest angular 
mode $\Psi_0^{\mr{\theta}}(\mb{\hat{r}})$. To simplify the calculations, we assumed 
$\Psi_0^{\mr{\theta}}(\mb{\hat{r}})=1$. To select large-scale modes, we adopted a power-law $P_0(k)$
so the signal is largest on the largest scales. The radial KL modes form an orthonormal basis 
in the redshift interval probed by the survey ($0<z<2$).

\section{Building pKL modes for non-separable selection functions}
\label{sec:nonsep-pKL}

Despite the efforts made by imaging and spectroscopic projects to cover the sky in a homogeneous 
way, it is very difficult to accomplish such task for the whole sky. For instance, on top of 
unanticipated calibration and technical problems and changes in pipelines that affected different 
regions of the sky, the SDSS data presented small differences between the north and south Galactic 
hemispheres \cite{Reid16}. Moreover, different instruments (such as those of the Euclid's ground 
segment) also tend to result in slightly different galaxy selection functions and contamination rates. 
Given these challenges, we extended the pKL method for survey conditions that are non-separable into radial and 
angular parts; as a bonus, this extension allows the combined analysis of multiple surveys.   

\begin{figure}
  \center
  \includegraphics[width=0.52\textwidth]{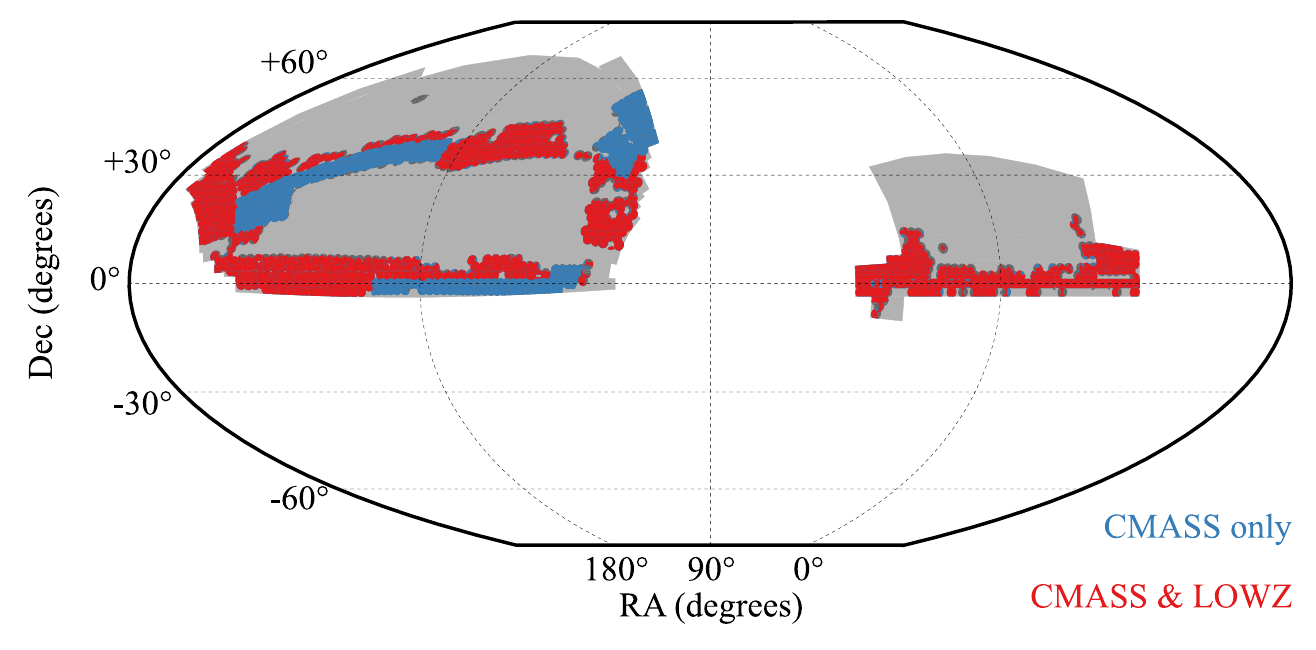}
  \includegraphics[width=0.46\textwidth]{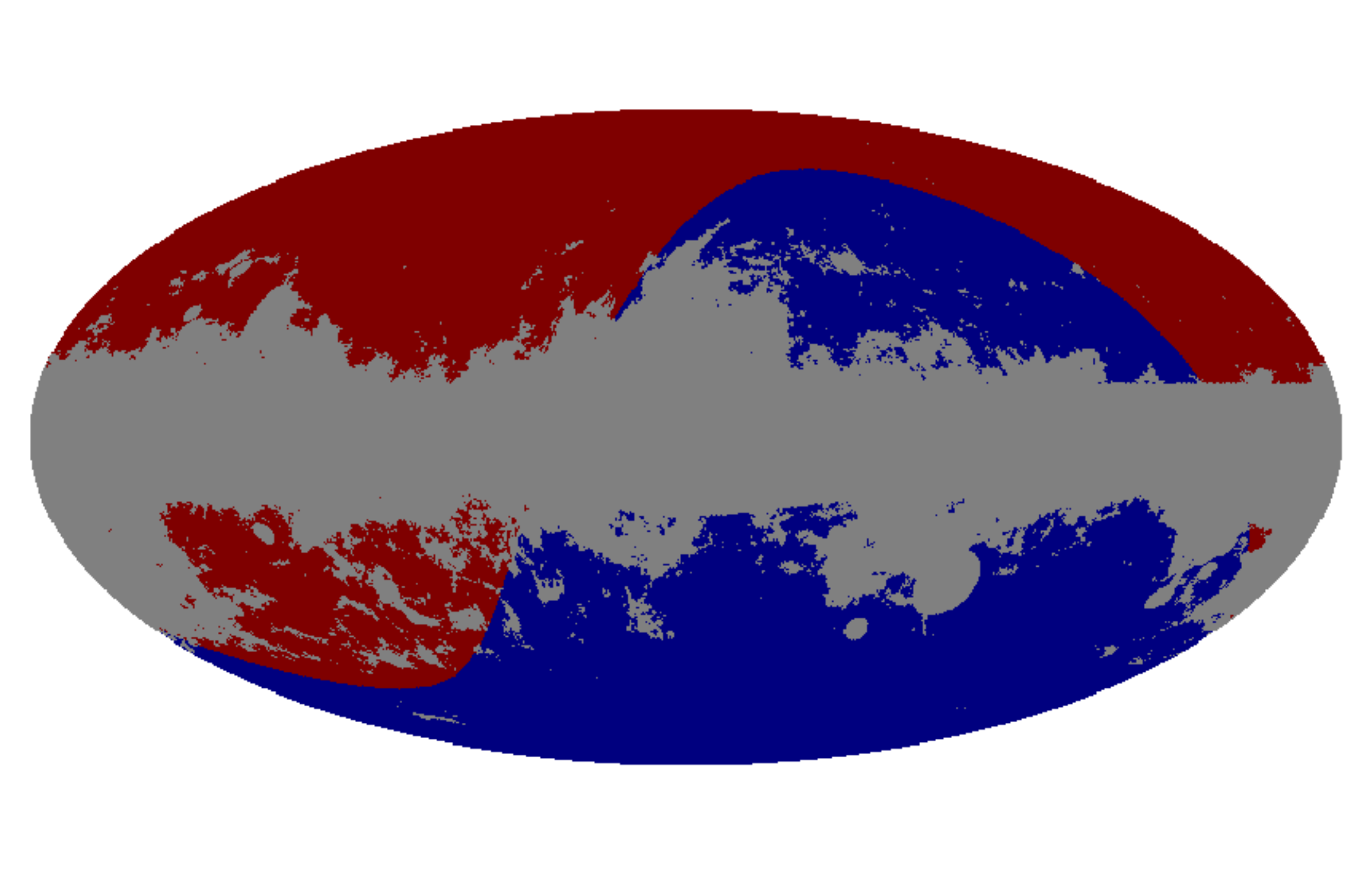}
  \caption{\emph{Left panel:} map of the SDSS footprint in equatorial coordinates. The final BOSS footprint, with disjoint 
North (on the left) and South (on the right) Galactic Caps that present slightly different radial selection 
functions, is shown in gray. The colored dots represent the DR9 footprint: regions only containing 
observations of galaxies from the CMASS sample are presented in blue, while regions that also contain 
observations of galaxies from the LOWZ sample are presented in red. These two samples have very different 
radial selection functions. Figure from \cite{Ross12b}. \emph{Right panel:} mask of the 
WISE$\times$SuperCOSMOS catalog, in galactic coordinates, with the masked regions in gray. The red region 
is the northern equatorial hemisphere, observed with the POSS-II telescope, while the blue region is the 
southern equatorial hemisphere, observed with the UKST telescope \cite{Bilicki16}.}
  \label{fig:sdss-footprint}
\end{figure}

Our approach to deal with non-separable conditions is to assume they can be defined on 
angular sub-domains (i.e. they are piece-wise functions) and that in each sub-domain the separation between 
radial and angular parts is valid. This approach is adequate for the SDSS and WISE$\times$SuperCOSMOS 
\cite{Bilicki16} datasets, and should be valid for Euclid and LSST and also for combining data from 
different surveys. A concrete example is the analysis of SDSS Data Release 9 (DR9) galaxy distribution (see the 
left panel of Fig. 
\ref{fig:sdss-footprint}), where we could describe the galaxy selection function in three sub-domains: the 
South Galactic Cap (SGC), and the regions in the North Galactic Cap (NGC) containing CMASS only and 
CMASS \& LOWZ galaxies. In each of these sub-domains, the selection function can be considered separable. 
Since the window function $W(\mb{r})$ multiplies all terms in Eq. \ref{eq:nobs-def}, it can be used as 
a ``switch'' for each one of the different separable selection functions defined over each sub-domain 
$h$: $W(\mb{r}) = \sum_h W_h(\mb{r})$, where the non-zero regions of $W_h(\mb{r})$ do not overlap. 
Another real example is the WISE$\times$SuperCOSMOS catalog, built from observations made with 
different telescopes that lead to differences between the two equatorial hemispheres \cite{Bilicki16,Xavier19}. 
These hemispheres can be described as sub-domains of separable selection functions (see the right panel of 
Fig. \ref{fig:sdss-footprint}).

For simplicity, we treat the case where there are two sub-domains with different $\bar{n}_{\mr{obs}}(\mb{r})$, 
indicated by the indices N and S (as in `North' and `South'); however, the treatment of an arbitrary 
number of sub-domains is exactly the same and the generalization straightforward. In mathematical terms, 
this is given by: 

\begin{equation}
\begin{split}
n_{\mr{obs}}(\mb{r}) = 
&W_{\mr{\theta}}^{\mr{N}}(\mb{\hat{r}}) W_{\mr{z}}^{\mr{N}}(r) \{ \bar{n}_{\mr{g}}^{\mr{\theta}}(\mb{\hat{r}}) \bar{n}_{\mr{g}}^{\mr{z,N}}(r) 
[1+\delta_{\mr{g}}^{\mr{N}}(\mb{r})] + \epsilon(\mb{r}) + s_{\mr{\theta}}^{\mr{N}}(\mb{\hat{r}}) s_{\mr{z}}^{\mr{N}}(r) \} + \\
&W_{\mr{\theta}}^{\mr{S}}(\mb{\hat{r}}) W_{\mr{z}}^{\mr{S}}(r) \{ \bar{n}_{\mr{g}}^{\mr{\theta}}(\mb{\hat{r}}) \bar{n}_{\mr{g}}^{\mr{z,S}}(r) 
[1+\delta_{\mr{g}}^{\mr{S}}(\mb{r})] + \epsilon(\mb{r}) + s_{\mr{\theta}}^{\mr{S}}(\mb{\hat{r}}) s_{\mr{z}}^{\mr{S}}(r) \}.
\end{split}
\label{eq:ns-nobs-def}
\end{equation}
In the equation above, we did not assign different angular parts $\bar{n}_{\mr{\theta}}(\mb{\hat{r}})$ to each sub-domain 
because the different $\mb{\hat{r}}$ already takes care of that [the same will happen to angular weights 
$w_{\mr{\theta}}(\mb{\hat{r}})$]. It is also worth remembering that 
the window function for one sub-domain is zero over the other sub-domains, and thus 
$W_{\mr{\theta}}^{\mr{N}}(\mb{\hat{r}})W_{\mr{\theta}}^{\mr{S}}(\mb{\hat{r}})=0$ for all $\mb{\hat{r}}$.
We also considered the possibility that galaxy bias might be different in each sub-domain (e.g. due to 
different selection criteria). Luckily, the galaxy bias is always multiplied, in the equations, by the 
angular window function, such that in each sub-domain the bias can be described as dependent only on 
$r$. 

\subsection{Computing the angular pKL modes}
\label{sec:ns-ang-mechanics}

We proceed through the same method described in Secs. \ref{sec:kl-mechanics} and \ref{sec:pseudo-kl}, 
by computing the optimal angular modes. The first step is to determine the $\M{\Pi}$ matrix.
Since the systematics $M_j(\mb{r})$ describe $\langle n_{\mr{obs}}(\mb{r}) \rangle$, they will also be described 
as piece-wise functions, e.g.: $M_j(\mb{r})=M_j^{\mr{N}}(\mb{r})+M_j^{\mr{S}}(\mb{r})$. From the definition of 
$M_j(\mb{r})$, we see this corresponds to simply increasing the number of systematic templates while the 
process of computing $\M{\Pi}$ remains the same.

We continue by computing the angular modes' pre-whitening matrix from $\M{N'}$. 
Since the Poisson noise $\epsilon(\mb{r})$ does not correlate at non-zero distances and  
$W_{\mr{\theta}}^{\mr{N}}(\mb{\hat{r}})$ and $W_{\mr{\theta}}^{\mr{S}}(\mb{\hat{r}})$ do not overlap, the basis modes' 
noise covariance matrix $\M{N'}$ is simply the sum of the contributions coming from different sub-domains, 
each one computable by Eq. \ref{eq:phi-noise-sep}. Therefore, $\M{N'}$ becomes non-separable. 
In case $W_{\mr{z}}^{\mr{N}}(r) = W_{\mr{z}}^{\mr{S}}(r)$, this complication can be averted if and we adopt the 
radial weights $w_{\mr{z},h}(r) = r \sqrt{\bar{n}_{\mr{obs}}^{\mr{z,h}}(r)}$ ($h=\mr{N,S}$), thus making the 
radial part of $\M{N'}$ the same for every sub-domain and allowing us to factor it out. These weights 
happen to be the same as those that optimize the angular pKL modes (see Sec. \ref{sec:pkl-ang}).
In any case, the noise covariance matrix of the angular basis functions is computed from Eq. \ref{eq:phi-noise-sep} 
by fixing $\Phi_i^{\mr{z}}(r)=1$. Explicitly, we have:

\begin{equation}
N'_{ij} = \int \left[ 
  I_{\mr{N}} W_{\mr{\theta}}^{\mr{N}}(\mb{\hat{r}}) \frac{\bar{n}_{\mr{obs}}^{\mr{\theta}}(\mb{\hat{r}})}{w_{\mr{\theta}}^2(\mb{\hat{r}})} + 
  I_{\mr{S}} W_{\mr{\theta}}^{\mr{S}}(\mb{\hat{r}}) \frac{\bar{n}_{\mr{obs}}^{\mr{\theta}}(\mb{\hat{r}})}{w_{\mr{\theta}}^2(\mb{\hat{r}})} \right] 
\Phi_i^{\mr{\theta}}(\mb{\hat{r}}) \Phi_j^{\mr{\theta}*}(\mb{\hat{r}}) \mr{d^2}\hat{r},
\label{eq:piecewise-noise-cov}
\end{equation}
\begin{equation}
I_h \equiv \int W_{\mr{z}}^{\mr{h}}(r) \frac{\bar{n}^{\mr{z,h}}_{\mr{obs}}(r)}{w_{\mr{z},h}^2(r)} r^2\mr{d}r.
\label{eq:i-def}
\end{equation}
If we use $\Phi_i^{\mr{\theta}}(\mb{\hat{r}})=Y^*_{\ell_i m_i}(\mb{\hat{r}})$, then:
\begin{equation}
\begin{split}
N'_{ij} = & J_{\ell_i m_i \ell_j m_j}\left[ 
  I_{\mr{N}} W_{\mr{\theta}}^{\mr{N}} \frac{\bar{n}_{\mr{obs}}^{\mr{\theta}}}{w_{\mr{\theta}}^2} + 
  I_{\mr{S}} W_{\mr{\theta}}^{\mr{S}} \frac{\bar{n}_{\mr{obs}}^{\mr{\theta}}}{w_{\mr{\theta}}^2} \right] \\ = &  
  I_{\mr{N}} J_{\ell_i m_i \ell_j m_j}\left[ 
  W_{\mr{\theta}}^{\mr{N}} \frac{\bar{n}_{\mr{obs}}^{\mr{\theta}}}{w_{\mr{\theta}}^2} \right] + 
  I_{\mr{S}} J_{\ell_i m_i \ell_j m_j}\left[
    W_{\mr{\theta}}^{\mr{S}} \frac{\bar{n}_{\mr{obs}}^{\mr{\theta}}}{w_{\mr{\theta}}^2} \right].
\end{split}
\label{eq:piecewise-noise-Jlmlm}
\end{equation}

To derive the signal covariance matrix $\M{S'}$, 
we need to integrate the (weighted) observed density (Eq. \ref{eq:ns-nobs-def}) along the line of sight to work 
with the projected density contrast $\sigma_{\mr{g}}(\mb{\hat{r}})$. In this process, 
different selection functions will lead to different projected densities, so:
$W_{\mr{\theta}}(\mb{\hat{r}}) \sigma_{\mr{g}}(\mb{\hat{r}}) = W_{\mr{\theta}}^{\mr{N}}(\mb{\hat{r}}) \sigma_{\mr{g}}^{\mr{N}}(\mb{\hat{r}}) + W_{\mr{\theta}}^{\mr{S}}(\mb{\hat{r}}) \sigma_{\mr{g}}^{\mr{S}}(\mb{\hat{r}})$, where $\sigma_{\mr{g}}^{\mr{h}}(\mb{\hat{r}})$ 
is computed from Eq. \ref{eq:proj-density} but with all functions specified for the sub-domain $h$.
From Eq. \ref{eq:ang-signal-1}, we see that $\M{S'}$ will be, like $\M{N'}$, a sum of contributions from
different sub-domains; however, unlike $\M{N'}$, the signal from different sub-domains are correlated. 
Thus, we have:

\begin{equation}
S'_{ij} = {S'}^{\mr{NN}}_{ij} + {S'}^{\mr{NS}}_{ij} + {S'}^{\mr{SN}}_{ij} + {S'}^{\mr{SS}}_{ij},
\label{eq:ns-signal-sum}
\end{equation}
where (in a similar way as in Eq. \ref{eq:phi-ang-signal}):

\begin{equation}
{S'}^{hk}_{ij} \equiv 
J_{[\ell_i m_i][\ell m]}\left[\frac{W_{\mr{\theta}}^{\mr{h}} \bar{n}^{\mr{\theta}}_{\mr{g}}}{w_{\mr{\theta}}}\right]
C_{[\ell m][\ell' m']}^{hk}
J^\dagger_{[\ell' m'][\ell_j m_j]}\left[\frac{ W_{\mr{\theta}}^{\mr{k}}\bar{n}^{\mr{\theta}}_{\mr{g}}}{w_{\mr{\theta}}}\right].
\end{equation}
In the equation above, $C_{[\ell m][\ell' m']}^{hk}=C^{hk}_{(\ell)}\delta^{\mr{K}}_{\ell\ell'}\delta^{\mr{K}}_{mm'}$ 
is the cross angular power spectrum (full-sky) of the projected densities 
$\sigma_{\mr{g}}^{h}(\mb{\hat{r}})$ and $\sigma_{\mr{g}}^{k}(\mb{\hat{r}})$.

Eq. \ref{eq:ns-signal-sum} shows an interesting feature of the data: in principle, there is information 
in the cross-correlation between the two disjoint sectors (i.e. the total signal variance is not just the 
sum of the variances in each sector because the data has large scale correlations). To estimate the relevance 
of these cross terms, we considered the case of a hypothetical survey with the two sectors  
shown in the right panel of Fig. \ref{fig:sdss-footprint}, in red and in blue. To make things simple, we assumed the 
only potential difference between the two sectors is the mean projected density.

\begin{figure}
  \center
  \includegraphics[width=0.49\textwidth]{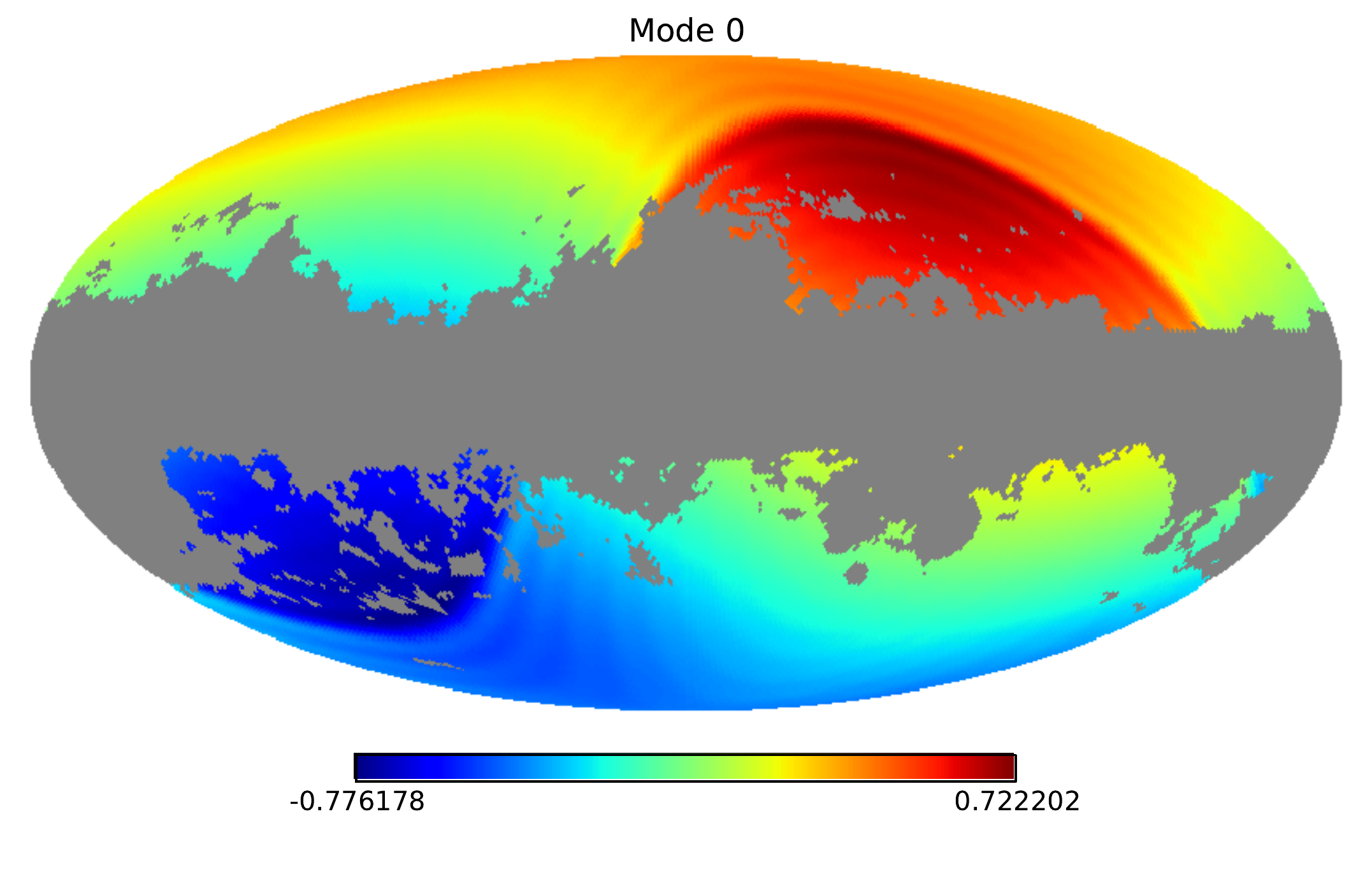}
  \includegraphics[width=0.49\textwidth]{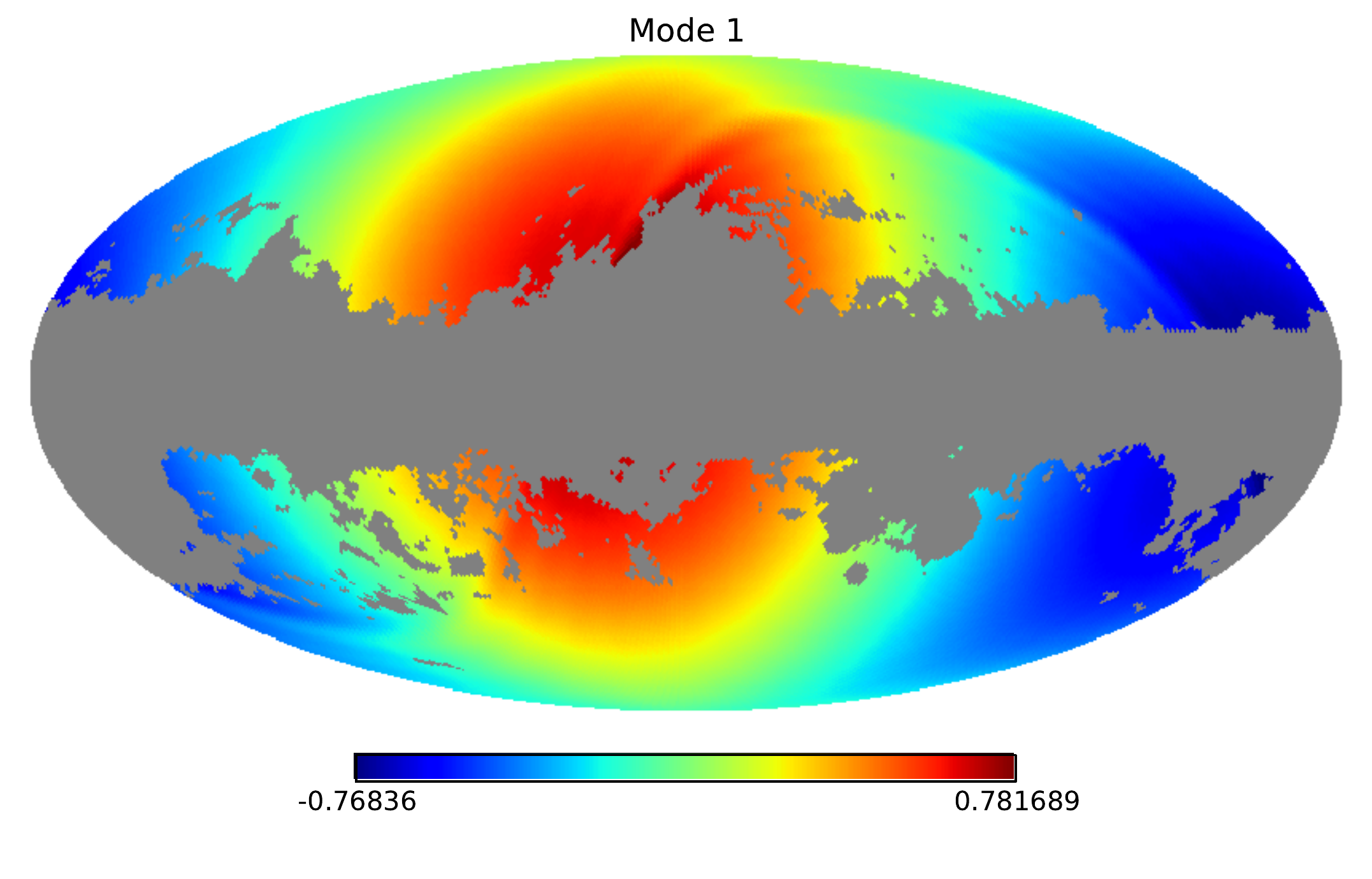}
  \includegraphics[width=0.49\textwidth]{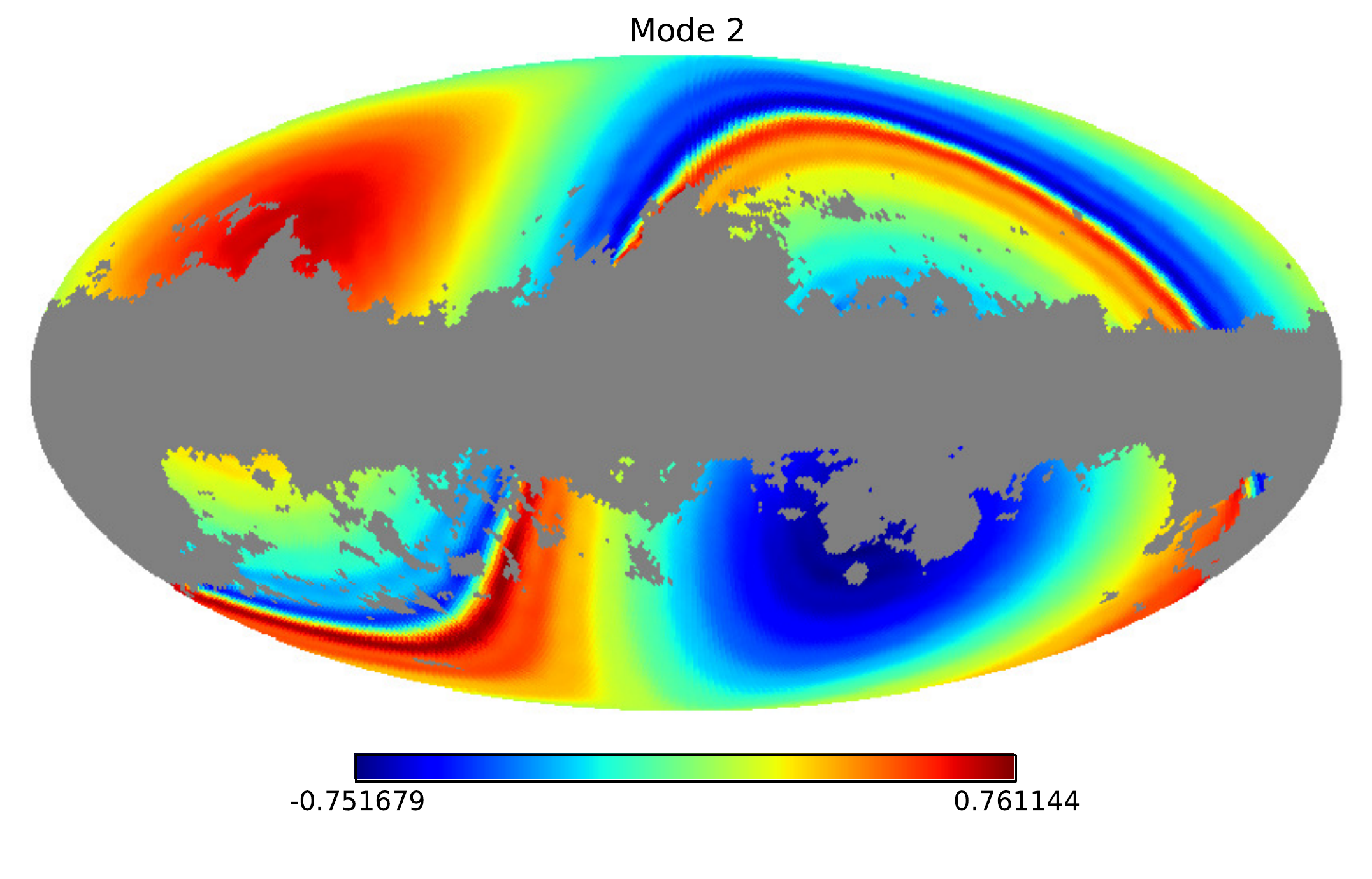}
  \includegraphics[width=0.49\textwidth]{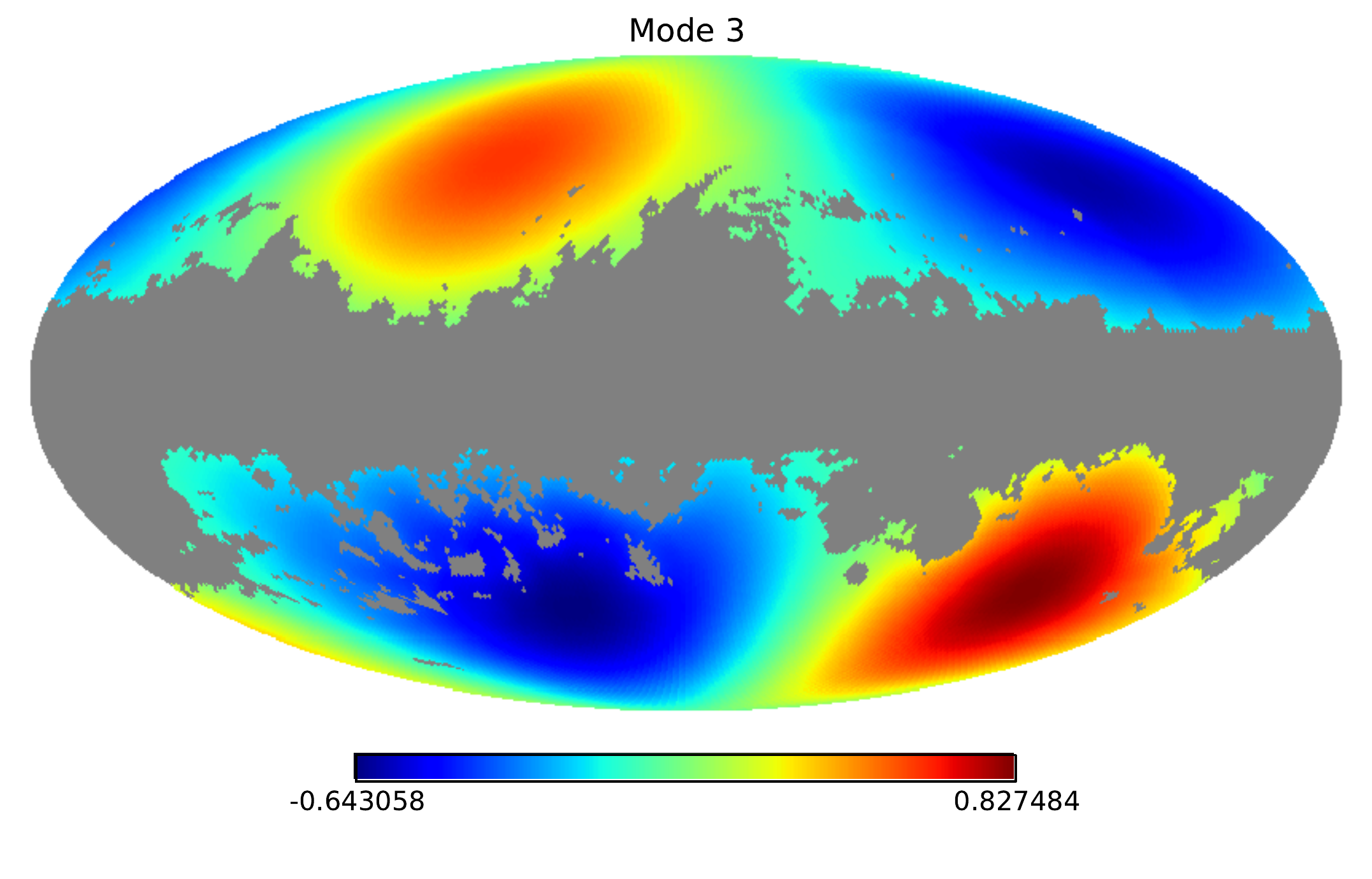}
  \caption{The first four angular KL modes to probe the largest scales, assuming a survey comprised of 
      two disjoint regions depicted in the right panel of Fig. \ref{fig:sdss-footprint}. The oscillations 
      near the border between the northern and southern hemispheres, easily seen in modes 0 and 2, are artifacts known 
      as Gibbs phenomenon.}
  \label{fig:wsc-modes}
\end{figure}

We employed a monotonically decreasing $C_\ell$ to enforce the building of angular KL modes that 
probe the largest scales. The first four derived angular KL modes are shown in Fig. \ref{fig:wsc-modes}. 
Specially from the first three modes, it is easy to see that they are all orthogonal to the mean density 
in each hemisphere, separately. That is, uncertainties on the mean density in each hemisphere (and thus on their 
difference) do not affect the measured mode amplitude. Secondly, we note that none of the modes probe each 
hemisphere individually; they all account for density fluctuations in both hemispheres at the same time. This 
evidences that there is information in the cross-correlation between both hemispheres. 

\begin{figure}
  \center
  \includegraphics[width=0.8\textwidth]{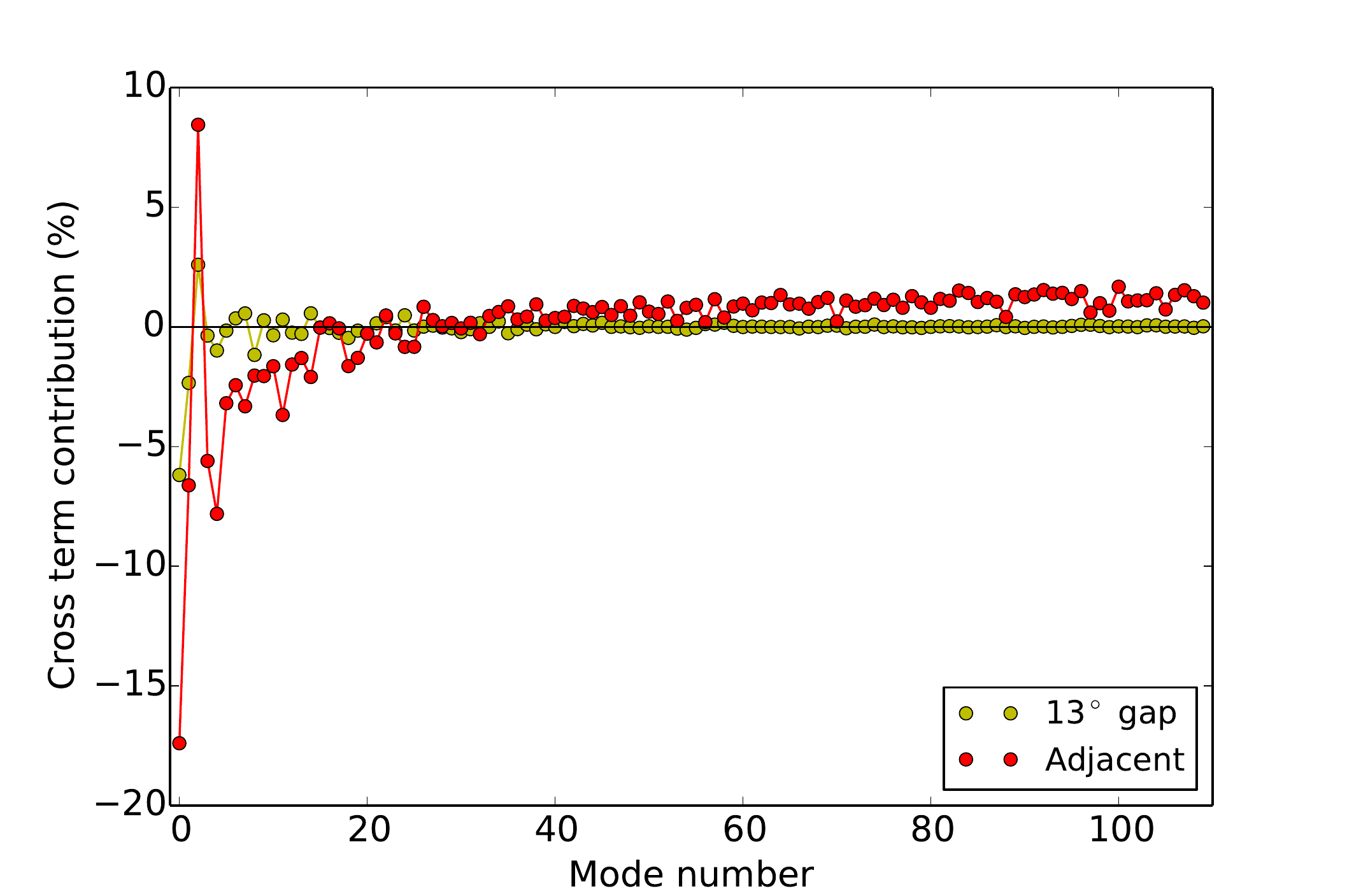}
  \caption{Fraction of the expected variance of the coefficients describing the weighted projected density 
      in terms of angular KL modes that comes from cross-correlations between two disjoint sky sectors, 
      as a function of mode number (largest scales first). The red dots represent the results obtained for the 
      two hemispheres shown in the right panel of Fig. \ref{fig:sdss-footprint}, while the ochre dots represent 
      the results for the same hemispheres but with a $13^\circ$ wide buffer zone (masked region) between the two hemispheres. 
      See the text for the exact expression generating these values.}
  \label{fig:x-contribution}
\end{figure}

We also can verify that the KL modes extract information across sectors by computing their fractional 
contribution to the diagonal of the total signal matrix, that is: 
$(\M{S}^{\mr{NS}}+\M{S}^{\mr{SN}})_{(ii)}/(\M{S}^{\mr{NN}}+\M{S}^{\mr{NS}}+\M{S}^{\mr{SN}}+\M{S}^{\mr{SS}})_{(ii)}$, where 
$\M{S}^{\mr{hk}} = \mathcal{M}\M{S'}^{\mr{hk}}\mathcal{M}^\dagger$ and $_{(ii)}$ denotes the $i$th element of the diagonal. 
Fig. \ref{fig:x-contribution} shows these quantities 
for the case discussed here (two sectors as in the right panel of Fig. \ref{fig:sdss-footprint}) and for the same case but with a 
13$^\circ$-wide buffer zone between the northern and southern hemispheres (i.e. we masked out the frontier 
between the two sectors). We see that the cross-terms contribute to the signal, specially on the largest scales where it reaches 
up to $\sim$15\% of the total. The fact that the buffer zone significantly reduces the cross-term contribution tells us 
that the information across sectors comes from data correlations near their border. This makes sense as density correlations 
rapidly decrease with distance but still do not care if they cross human-made boundaries.

\subsection{Computing the radial pKL modes}

Just like for the angular part described in the previous section, the separation of the selection 
function in sub-domains multiplies the amount of systematic templates the radial part has to deal with. 
That is, the matrix $\M{\Pi}$ needs to project out from $\Psi^{\mr{z}}_i(r)$ the radial part $M_j^{\mr{z}}(r)$ 
of all templates (from all sub-domains). 

When dealing with piece-wise separable selection functions like we do in this section, 
the process of pre-whitening the noise covariance matrix for the 3D pKL modes (Eq. \ref{eq:W-derivation}) 
can be slightly more complicated if the radial window functions for each sub-domain $h$, $W_{\mr{z}}^h(r)$, 
are different: in this case we cannot use radial weights to factor out the radial part 
(see Sec. \ref{sec:ns-ang-mechanics}). Once the angular pKL modes $\Psi_i^{\mr{\theta}}(\mb{\hat{r}})$ 
are defined from Eq. \ref{eq:piecewise-noise-cov}, the matrix $\M{N'}$ for the 3D pKL modes will 
be a sum of contributions from each sub-domain:

\begin{equation}
\begin{split}
N'_{[ni][mj]} = &
\int W_{\mr{z}}^{\mr{N}}(r) \frac{\bar{n}^{\mr{z,N}}_{\mr{obs}}(r)}{w^2_{\mr{z,N}}(r)} \Phi_n^{\mr{z}}(r) \Phi_m^{\mr{z}*}(r) r^2\mr{d}r
\int W_{\mr{\theta}}^{\mr{N}}(\mb{\hat{r}}) \frac{\bar{n}^{\mr{\theta}}_{\mr{obs}}(\mb{\hat{r}})}{w^2_{\mr{\theta}}(\mb{\hat{r}})} 
\Psi_i^{\mr{\theta}}(\mb{\hat{r}}) \Psi_j^{\mr{\theta}*}(\mb{\hat{r}}) \mr{d^2}\hat{r}~ + \\ &
\int W_{\mr{z}}^{\mr{S}}(r) \frac{\bar{n}^{\mr{z,S}}_{\mr{obs}}(r)}{w^2_{\mr{z,S}}(r)} \Phi_n^{\mr{z}}(r) \Phi_m^{\mr{z}*}(r) r^2\mr{d}r
\int W_{\mr{\theta}}^{\mr{S}}(\mb{\hat{r}}) \frac{\bar{n}^{\mr{\theta}}_{\mr{obs}}(\mb{\hat{r}})}{w^2_{\mr{\theta}}(\mb{\hat{r}})} 
\Psi_i^{\mr{\theta}}(\mb{\hat{r}}) \Psi_j^{\mr{\theta}*}(\mb{\hat{r}}) \mr{d^2}\hat{r}.
\end{split}
\label{eq:ns-3d-noise}
\end{equation}

The problem is that, as Eq. \ref{eq:piecewise-noise-cov} shows, $\Psi_i^{\mr{\theta}}(\mb{\hat{r}})$ are built 
to yield a unit noise covariance matrix only when applied to the combination of sub-domains N and S, weighted by 
$I_\mr{N}$ and $I_\mr{S}$ (Eq. \ref{eq:i-def}), whereas Eq. \ref{eq:ns-3d-noise} shows that a combination of 
radial basis modes $\Phi_n^{\mr{z}}(r)$ would only result in an unit 3D noise covariance matrix if 
$\Psi_i^{\mr{\theta}}(\mb{\hat{r}})$ led to a unit noise covariance matrix in each sub-domain separately. 
Consequently, we assume $W_{\mr{z}}^{\mr{N}}(r) = W_{\mr{z}}^{\mr{S}}(r) = W_{\mr{z}}(r)$ and adopt the weights:

\begin{equation}
w_{\mr{z,h}}(r)=r\sqrt{\frac{\bar{n}^{\mr{z},h}_{\mr{obs}}(r)}{I_h}},
\label{eq:piecewise-weights}
\end{equation}
allowing us to factor out the radial part:

\begin{equation}
\begin{split}
N'_{[ni][mj]} = &
\int W_{\mr{z}}(r) \Phi_n^{\mr{z}}(r) \Phi_m^{\mr{z}*}(r) \mr{d}r
\int \left[ 
  I_{\mr{N}} W_{\mr{\theta}}^{\mr{N}}(\mb{\hat{r}}) \frac{\bar{n}_{\mr{obs}}^{\mr{\theta}}(\mb{\hat{r}})}{w_{\mr{\theta}}^2(\mb{\hat{r}})} + 
  I_{\mr{S}} W_{\mr{\theta}}^{\mr{S}}(\mb{\hat{r}}) \frac{\bar{n}_{\mr{obs}}^{\mr{\theta}}(\mb{\hat{r}})}{w_{\mr{\theta}}^2(\mb{\hat{r}})} \right] 
\Psi_i(\mb{\hat{r}}) \Psi_j^*(\mb{\hat{r}}) \mr{d^2}\hat{r} \\
= & \int W_{\mr{z}}(r) \Phi_n^{\mr{z}}(r) \Phi_m^{\mr{z}*}(r) \mr{d}r ~\delta^{\mr{K}}_{ij}.
\end{split}
\end{equation}
In other words, under these conditions we only need to find a pre-whitening matrix $\M{W}$ that turns 
$\Pi_{kn}\int W_{\mr{z}}(r) \Phi_n^{\mr{z}}(r) \Phi_m^{\mr{z}*}(r) \mr{d}r ~\Pi^{\dagger}_{mp}$ into the 
identity matrix.

Finally, we need to compute a matrix $\M{S'}$ for the radial modes, one for each angular part $\Psi_i^{\mr{\theta}}(\mb{\hat{r}})$ 
(as in Sec. \ref{sec:pkl-radial}). Just like for the angular modes in the previous section, 
we can write $\M{S'}$ as a sum of contributions from each sub-domain (and from their cross-correlations). 
Explicitly, we obtain:

\begin{equation}
{S'}_{[in][jp]} = \frac{2}{\pi} \int P_0(k)\sum_{h,h'}\sum_{\ell, m}\tilde{G}_{\ell m}^{h,in}(k) \tilde{G}_{\ell m}^{h',jp*}(k)k^2\mr{d}k, 
\label{eq:ns-signal}
\end{equation}

\begin{equation}
\tilde{G}_{\ell m}^{h,in}(k) \equiv \mathcal{M}^\theta_{i[LM]}J_{[LM][(\ell m)]} 
\left[W_{\mr{\theta}}^{(h)}\frac{\bar{n}^{\mr{\theta}}_{\mr{g}}}{w_{\mr{\theta}}}\right] 
\int \Phi^{\mr{z}*}_n(r)\frac{\bar{n}_{\mr{g}}^{\mr{z}(h)}(r)}{w_{\mr{z}(h)}(r)}G^{(h)}_\ell(k,r)r^2\mr{d}r,
\end{equation}

\begin{equation}
G^h_\ell(k,r) \equiv b_h(r)D(r)j_\ell(kr) - \gamma_h(r)\frac{j'_\ell(kr)}{k}-D(r)f(r)j''_{\ell}(kr),
\end{equation}

\begin{equation}
\gamma_{h}(r) \equiv \frac{2D(r)f(r)}{r} + 
\frac{1}{\bar{n}^{\mr{z}(h)}_{\mr{g}}(r)}\frac{\partial[D(r)f(r)\bar{n}^{\mr{z}(h)}_{\mr{g}}(r)]}{\partial r}.
\end{equation}
In the equations above, the indices $i$ and $j$ identify the angular part of the pKL modes, while 
$n$ and $p$ identify the radial basis functions; $h$ and $h'$ informs to which sub-domain each function belongs to; and 
$\ell$ and $m$ are the usual spherical harmonic multipoles indices. We also remind that we adopt the Einstein 
summation convention except for indices inside parentheses, that compound indices (representing a single dimension) 
are written inside square brackets and that the weights $w_{\mr{z},h}(r)$ are given by Eq. \ref{eq:piecewise-weights}.
Again, the matrix $\M{K}$ is obtained for each angular mode $\Psi_i^{\mr{\theta}}(\mb{\hat{r}})$ following the third item in 
Sec. \ref{sec:kl-mechanics} and using Eq. \ref{eq:ns-signal} with $i=j$.

\section{Using pKL to measure $P_0(k)$}
\label{sec:apply-pkl}

The previous sections dealt with the issue of finding optimal (pKL) modes $\Psi_{i}(\mb{r})$ to 
measure the clustering of galaxies, given the survey's characteristics and assuming a fiducial $P_0(k)$. 
Although defining $\Psi_{i}(\mb{r})$ can be a lengthy process, once this is done it is straightforward to use 
them to measure $P_0(k)$.

The first step is to compute the coefficients $x_i$ that describe the observed density in terms of the pKL modes 
(Eq. \ref{eq:pkl-coeff}), using the radial weights given by Eq. \ref{eq:piecewise-weights}. As Eq. \ref{eq:x-mean} 
shows, these coefficients have mean value zero, a feature that is completely independent of cosmology and 
therefore remains no matter the true value of $P_0(k)$. The covariance can be described as 
$\langle x_i x_j^*\rangle = S_{ij} + \delta^{\mr{K}}_{ij}$, and this fact also does not depend on cosmology (the 
noise covariance matrix, which is the identity matrix, only depends on the characteristics of the survey). Thus, all cosmological information 
is encoded in the signal covariance matrix, and in a rather simple way:

\begin{equation}
S_{ij} = \int P_0(k) H_{ij}(k) \mr{d}k,
\label{eq:s-cov-measure}
\end{equation}
where $H_{ij}(k)$ is pre-computed once and for all according to the designed optimal modes:

\begin{equation}
H_{[ir][js]}(k) \equiv \frac{2}{\pi} \mathcal{M}^{(i)}_{rn} 
\sum_{h,h'}\sum_{\ell,m} \tilde{G}^{h,in}_{\ell m}(k) \tilde{G}^{h',jp*}_{\ell m}(k)k^2
\mathcal{M}^{(j)\dagger}_{ps}.
\end{equation}
Since $x_i$ follow a Gaussian distribution \cite{Tegmark04}, their likelihood function is well 
determined:
 
\begin{equation}
\mathcal{L} = \frac{1}{(2\pi)^{N/2}\sqrt{\det(\M{S}+\M{I})}}\exp\left[ -\frac{1}{2}\M{x}^\dagger(\M{S}+\M{I})^{-1}\M{x} \right],
\label{eq:likelihood}
\end{equation}
and we can take advantage of pKL compression and smaller number of modes that probe the largest scales to analyze 
$\mathcal{L}$ and find $P_0(k)$.

As a last strategy to speed up computations, we can model $P_0(k)$ as a piece-wise function of constant band-powers $p_q$:

\begin{equation}
P_0(k) = \sum_q p_q B_q(k),
\end{equation}
where $B_q(k)$ are rectangular functions. Under this approach, Eq. \ref{eq:s-cov-measure} can be written as 
a weighted sum of pre-computed matrices:

\begin{equation}
S_{ij} = \sum_q p_q\int  B_q(k) H_{ij}(k) \mr{d}k.
\end{equation}

\begin{figure}
  \center
  \includegraphics[width=0.7\textwidth]{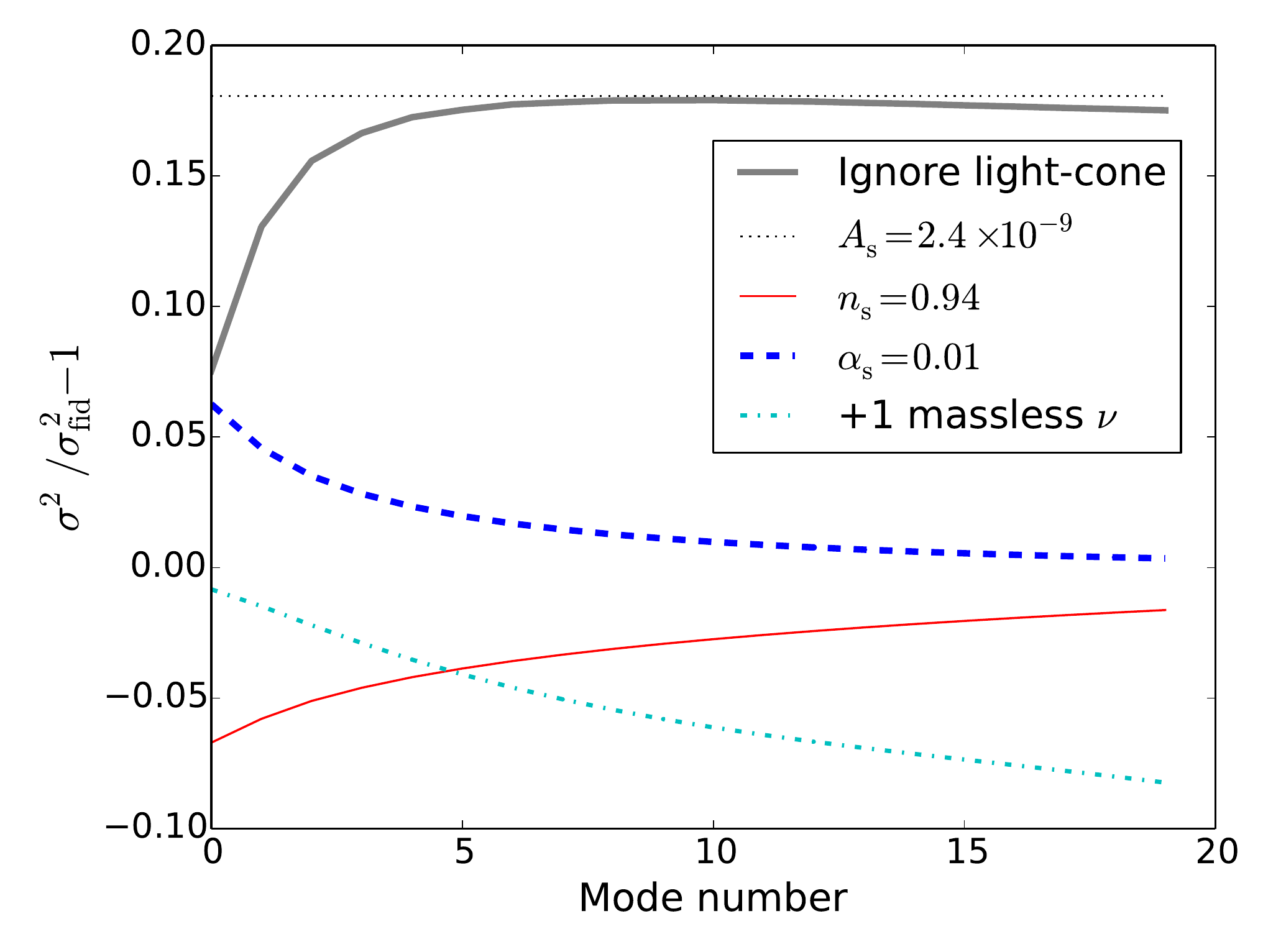}
  \caption{Fractional difference between the expected variance $\sigma^2$ of large-scale pKL modes 
      (see Fig. \ref{fig:radial-kl-modes}) computed for variations over cosmological parameters and the 
      variance  $\sigma^2_{\mr{fid}}$ computed for 
      a fiducial flat $\Lambda$CDM cosmology ($A_{\mr{s}}=2.2\times10^{-9}$, $n_{\mr{s}}=0.96$, $\alpha_{\mr{s}}=0$, 
      $\Omega_{\mr{b}}h^2=0.022$, $\Omega_{\mr{cdm}}h^2=0.12$ and 3 massless neutrinos). For each curve, we 
      only varied one parameter. We also show as a thick gray curve the variance calculated when neglecting 
      light-cone effects.}
  \label{fig:kl-variance-change}
\end{figure}

To highlight the importance of taking into account light-cone effects when measuring density fluctuations 
on the largest scales, we computed the signal covariance matrix $\M{S}$ for the pKL modes presented in Fig. 
\ref{fig:radial-kl-modes} assuming different cosmologies and compared to the case where light-cone effects 
are ignored. Fig. \ref{fig:kl-variance-change} shows the fractional difference of pKL modes' variances with respect 
to a fiducial cosmology for these alternative cosmologies or approach.

It is evident that neglecting light-cone 
effects produces changes in $\M{S}$ that are of the same order as reasonable changes in large-scale cosmological 
parameters (such as the spectral index $n_{\mr{s}}$ and running $\alpha_{\mr{s}}$ of the primordial power spectrum). 
Similar results might be expected if one is interested in measuring scale-dependent biases like those in 
\cite{Durrer03,Dalal08}. 
On smaller scales (larger mode numbers), the outcome of leaving light-cone effects out is degenerate to a constant 
factor change in $P_0(k)$ (i.e. changing the amplitude $A_{\mr{s}}$), and therefore such effects may not have to be 
taken into account at these scales.\footnote{Neglecting light-cone effects also changes the covariances, so these 
two responses might be disentangled.}  

\section{Discussion}
\label{sec:discussion}

In this paper we have presented in detail the pseudo Karhunen-Lo\`{e}ve (pKL) method and its application 
to cosmology. This method includes the modes' orthogonalization to systematics templates, a feature that 
makes the analysis less sensitive to contaminants and uncertainties on the mean density and/or 
observed dipole caused by our peculiar motion. This orthogonalization method has the advantage that it does 
not require the systematics contribution level to be determined; it only requires that systematics are 
properly modeled up to a constant factor. It is worth pointing out that this orthogonalization approach 
only mitigates additive systematics and not multiplicative ones. Strategies for dealing with multiplicative 
effects are given in \cite{Ross12b,Shafer15,Xavier19}.

During the orthogonalization process, the dimensionality of the space covered by the KL modes is 
reduced: all data is projected onto a sub-space that is orthogonal to the mean density and systematics 
templates. This hinders the subsequent step of pre-whitening the data because the covariance matrices 
become singular. We presented a solution for this in Eq. \ref{eq:w-def}, where we identify 
redundant modes by their null variance. It is worth pointing out that the original dimensionality can 
be restored by including the systematics templates as ``special modes'', since these are orthogonal to 
all other modes by construction. While the variance measured in these modes cannot be used to extract 
cosmological information, their amplitude provides an estimate of the level of contamination associated to 
a particular template.

As shown in Sec. \ref{sec:pseudo-kl}, we simplify the application of the KL method 
to 3D data by first creating close to optimal angular modes that are later combined with radial ones to 
create full 3D modes. We proposed that these angular modes are obtained by applying the KL method to 
weighted instead of unweighted projected density of galaxies. The appropriate weights are those that 
make the noise level the same in every redshift slice (see Eq. \ref{eq:piecewise-weights}).
Sec. \ref{sec:pseudo-kl} can also be applied to tomographic analysis of galaxy surveys, in
which case no simplifications to the KL method are made.

The SDSS and WISE$\times$SuperCOSMOS have indicated that achieving a constant radial selection 
function across the whole sky might be difficult, and this issue can affect future surveys such 
as the Euclid ground segment. Thus, we presented a generalization of previous implementations of the 
pKL method that can tackle this situation by segmenting the sky into patches with locally constant 
radial selection functions. This method can also be used to combine information from multiple surveys. 
Interestingly, in Sec. \ref{sec:ns-ang-mechanics} we have also demonstrated that such combination 
leads to more information that the sum of its parts, given that galaxy densities are correlated over 
large distances (and across survey boundaries). This property is not unique to the pKL method and can be 
demonstrated for pseudo angular power spectra analysis (p$C_\ell$) \cite{Alonso18} and 
the Landy-Szalay \cite{LandySzalay93} configuration space estimator.

In the derivation and analysis of radial modes, we have taken into account the fact that 
our observations are made on the surface of our past light cone, and thus the farther we 
look in space, the farther we look in time. For deep enough surveys, the growth of structure and 
galaxy bias evolution along our line of sight must be taken into account. As shown in Fig. 
\ref{fig:kl-variance-change}, these light-cone effects distort the observed clustering on the 
largest scales and could be confused (if not accounted for) with other physical effects like 
scale-dependent biases, different primordial spectral indices and their runnings. 
The combination of proper modeling of this evolution in the radial direction with the 
use of distinct sky sectors and their resulting synergy makes the method presented here 
a powerful tool to probe the largest scales in the Universe. 

\section*{Acknowledgments}

We thank Prof. Andrew Hamilton for clarifying concepts and methods related to this work and 
Prof. Michael Strauss for helpful discussions and feedbacks. The author was financially supported by 
FAPESP Brazilian funding agency.

\bibliographystyle{JHEP}
\bibliography{main}

\end{document}